\documentclass[11pt]{article}

\usepackage{arxiv}

\usepackage[utf8]{inputenc} 
\usepackage[T1]{fontenc}    
\usepackage{caption}
\usepackage{graphicx}
\usepackage{natbib}
\usepackage{url} %
\usepackage{lmodern}
\usepackage{xcolor}
\usepackage{xspace}
\usepackage{colortbl} %
\usepackage{soul, booktabs, multirow} %
\usepackage{float} %
\usepackage{bm}
\usepackage{commands}
\usepackage{amsthm,amsmath,amsfonts,amssymb}
\usepackage{algorithm, algorithmic}
\usepackage{hyperref}

\theoremstyle{plain}%
\newtheorem{theorem}{Theorem}%
\newtheorem{proposition}[theorem]{Proposition}%

\theoremstyle{remark}%
\newtheorem{assumption}{Assumption}[section]
\theoremstyle{definition}%
\newtheorem{definition}{Definition}

\newcommand{\hlf}[1]{\cellcolor{yellow}{#1}} %
\newcommand{\hls}[1]{\cellcolor{green}{#1}} %

\title{A L-infinity Norm Counterfactual and Synthetic Control Approach}

\author{
 Le Wang \\
  AAEC\\
 Virginia Tech, USA\\
  \texttt{lewangecon@vt.edu} \\
   \And
 Xin Xing \\
  Department of Statistics\\
  Virginia Tech, USA\\
  \texttt{xinxing@vt.edu} \\
  \And
 Youhui Ye \\
  Department of Statistics \\
  Virginia Tech, USA\\
  \texttt{yye1997@vt.edu} \\
}

\renewcommand{\today}{}  
\date{}

\begin{document}

\maketitle

\begin{abstract}
This paper reinterprets the Synthetic Control (SC) framework through the lens of weighting philosophy, arguing that the contrast between traditional SC and Difference-in-Differences (DID) reflects two distinct modeling mindsets: \emph{sparse} versus \emph{dense} weighting schemes. Rather than viewing sparsity as inherently superior, we treat it as a modeling choice -- simple but potentially fragile. We propose an $L_{\infty}$-regularized SC method that combines the strengths of both approaches. Like DID, it employs a denser weighting scheme that distributes weights more evenly across control units, enhancing robustness and reducing overreliance on a few control units. Like traditional SC, it remains flexible and data-driven, increasing the likelihood of satisfying the parallel trends assumption while preserving interpretability. We develop an interior point algorithm for efficient computation, derive asymptotic theory under weak dependence, and demonstrate strong finite-sample performance through simulations and real-world applications.
\end{abstract}


\section{Introduction}

A central goal across disciplines such as economics, social sciences, and business is to uncover causal relationships -- to identify the true cause of an outcome and evaluate the impact of policies or interventions. Achieving this goal is inherently challenging due to the fundamental problem of causal inference: the impossibility of observing counterfactual outcomes; that is, outcomes that would have occurred under an alternative policy, action, or treatment.

Many widely used causal inference techniques can be understood as different strategies for approximating these unobservable counterfactuals using untreated (or control) units. Among them, the synthetic control (SC) method, introduced by \cite{abadie_synthetic_2010, abadie_comparative_2015}, offers an elegant and formal approach. It constructs a ``synthetic'' control group by assigning optimized weights to a combination of control units so that their weighted average closely mirrors the pre-treatment outcomes of the treated unit. The resulting synthetic control mimics the treated unit’s historical trajectory, enabling a credible estimate of what would have happened in the absence of the intervention. The intuitive and flexible nature of the SC method has made it a widely adopted tool for policy evaluation across disciplines. As \cite{athey_state_2017} observed, it represents “arguably the most important innovation in the policy evaluation literature in the last 15 years.”\footnote{The application of the SC approach has rapidly expanded (e.g., \cite{billmeier2013assessing} in economics, \cite{rehkopf2018new} in Epidemiology, and \cite{tirunillai2017does} in marketing). Furthermore, recent advances in causal inference have also leveraged the SC method for heterogeneous treatment effect estimation in the panel data framework \citep{shen2022heterogeneous}.}

Although it differs in implementation from the Difference-in-Differences (DID) method, SC implicitly relies on its own version of the parallel trends assumption—namely, that the treated unit and its synthetic counterpart would have followed parallel trajectories in the absence of treatment \citep{li_statistical_2020} (see also Section \ref{sec:method}). That is, the close pre-treatment alignment in the trajectories between the treated unit and its \emph{synthetic control} is assumed to persist into the post-treatment period had the treatment not occurred. This assumption is crucial for the credibility of the estimated counterfactual.

In this light, one of the fundamental differences between DID and SC lies rather in how they utilize information from the control units, or the weighting scheme that each method chooses to construct a counterfactual outcome. Specifically, DID applies a ``dense'' weighting scheme with equal weights to \emph{all} control units. In contrast, SC employs a ``sparse'' scheme to construct a weighted average of control units -- the ``synthetic control'' -- with weights chosen to closely match the pre-treatment outcome trajectory of the treated unit. When determining the optimal weights, the original SC imposes three key restrictions (and hence implicit assumptions): (1) the weights must sum to one, (2) they must be nonnegative, and (3) the model assumes a zero intercept. As shown in \cite{doudchenko_balancing_2016}, this approach is equivalent to regularization techniques such as Lasso, which inherently favors \emph{sparse} solutions, effectively selecting a small subset of control units to construct the counterfactual.

Importantly, we emphasize that sparsity versus denseness is not a question of correctness but one of modeling philosophy and practical trade-offs. Crucially, it should be noted that the sparsity feature of the SC is not the same as the \emph{purposeful} sparsity assumptions made in the machine learning literature. Instead, such feature arises as a mechanical byproduct of the optimization problem, and it is neither intentional nor necessarily advantageous. In fact, \cite{doudchenko_balancing_2016} argue that the underlying preference for concentrated weights or sparsity can be a controversial objective to formally justify. Furthermore, the sparsity feature may not hold if control units are considered to be drawn from a common distribution \citep{li_statistical_2020}.  Highly concentrated weights on a few control units can also lead to several practical challenges. First, such concentration may amplify an estimator's sensitivity to the specific characteristics of a few units, leading to higher variance and potentially unreliable results. Second, the reliance on a small subset of control units may introduce bias, as even minor deviations in these units can disproportionately affect the synthetic control estimate. 

More importantly, the transparency associated with the sparsity feature should not be conflated with interpretability. Sparse solutions can lead to counterintuitive weight distributions, driven by algorithmic constraints rather than substantive similarity. In Abadie's well-known application of synthetic control to the California Tobacco Control Program, for example, the sparsity outcome produced weights that is not straightforward or intuitive to understand: ``Utah and Nevada receive the largest weights while Connecticut, which might be culturally closer to California, receives the smallest. Plots of cigarette sales for those three states \dots make the relative weights even more difficult to understand.'' \citep{mcclelland2017synthetic}.

Our reinterpretation of varying approaches based on the underlying weighting schemes parallels recent advances in economic forecasting that have emphasized the value of complex models that rely on densely distributed coefficients to improve predictive performance. The intuition, as articulated in \citet{liao2023economic} and related work (\cite{shi2025ℓ}), lies in economic predictive signals being spread across many regressors. This allows for modest forecast bias while diversifying away overall variance -- even when many of the included predictors are pure noise. These insights support the view that, in many economic applications, dense weighting structures may offer more robust and reliable performance than sparse alternatives.

Here, we propose a novel SC estimator that leverages ``dense'' or ``denser'' weighting schemes based on the infinity norm regularization. The infinity norm, also known as the maximum norm or Chebyshev norm, has properties that allow us to develop a \emph{dense} weighting scheme. Similar to the synthetic DID approach (\cite{arkhangelsky2021synthetic}), our method serves as a middle ground between the traditional DID and SC approaches, combining the strengths of both while addressing their limitations. However, our solution is based on the SC framework, hence applicable in situations with a small number of treated units -- sometimes even a single treated unit, commonly encountered in policy evaluation contexts. On the one hand, unlike the uniform dense weighting scheme used in traditional DID, our approach imposes a flexible, data-driven weighting scheme that increases the likelihood of satisfying the parallel trends assumption. By assigning weights based on pre-treatment outcome similarity, our method reduces the risk of bias that arises when control units differ meaningfully in outcome dynamics, which ensures that the synthetic control more accurately reflects the untreated counterfactual. On the other hand, unlike sparse approaches (variants of the synthetic control approach), our method assigns weights to a broader set of control units, ensuring a more diverse synthetic control. Meanwhile, our approach constrains the largest weight and thus ensures that no individual control unit dominates the estimation. By reducing the reliance on a few units, our approach enhances the reliability of the synthetic control estimator while maintaining the desired level of match between the treated unit and the synthetic counterpart, as well as the same level of transparency (the weight on each unit is transparent). 

We derive the relationship between the proposed constrained estimator and the conventional least square estimator. Such results facilitate the comparisons between our estimator and the Lasso and other existing regularization methods. The comparisons elucidate a more complex penalization procedure that requires a more suitable numerical algorithm for solutions. As a result, we further demonstrate that the equivalence of our model to a quadratic constrained optimization problem, which can then be efficiently solved using the interior point method. 

We derive the asymptotic distribution of the proposed estimators under weak dependence conditions. As noted by \cite{li_statistical_2020}, despite the widespread adoption of the synthetic control (SC) method, there remains limited formal inference theory for estimating the average treatment effect (ATE) in long panel settings -- where both the pre- and post-treatment periods are substantial and the underlying conditions are general. Most existing inference procedures rely on placebo tests or permutation-based methods, which are typically calibrated for short post-treatment horizons and do not provide valid asymptotic guarantees in longer panels. This extends the work of \cite{li_statistical_2020}, who established asymptotic distributions for stationary, trend-stationary, and unit-root nonstationary cases under the assumption of exponential decay in correlation rates. Our framework provides a more comprehensive characterization that accommodates a broader class of dependency structures. 

We demonstrate the finite-sample performance through simulations. Such performances also provide benchmark for us to consider the discrepancies between varying methods in practice. Two real examples further illustrate the practical applicability and effectiveness of our methods and the concept of denser weighting schemes. In the first evaluation of the California Tobacco policy, we demonstrate that our proposed methods closely align with alternative approaches in achieving both a desirable match and an accurate estimation of treatment effects. This outcome is expected, as the tobacco market is relatively stable, and reliance on a small number of control units is less likely to introduce significant bias over time. In contrast, when analyzing a regulatory policy in the U.S. stock market—a more volatile environment subject to frequent shocks—overreliance on a limited set of control units can introduce substantial bias over time. Indeed, we find that conventional and Lasso-based estimates significantly underestimate both the treatment effect and the policy's effectiveness.

In Section~\ref{sec:method}, we present our methodology and demonstrate its properties through two propositions.
Section~\ref{sec:imp} introduces the Interior Point Method used to solve the optimization problem and outlines the algorithm.
In Section~\ref{sec:theory}, we reformulate the original question as a constrained optimization problem and leverage the theoretical framework of \cite{fu_2000_asymptotics} and \cite{li_statistical_2020} to derive the asymptotic distributional results.
Sections~\ref{sec:sim} and \ref{sec:real_data} compare the performance of our method to existing approaches, highlighting its advantages.
Finally, Section~\ref{sec:discussion} concludes with a brief discussion.

\section{Method}\label{sec:method}

We consider a panel data setting in which there are $J+1$ cross-sectional units observed in time periods $t = 1, \dots, T$. For ease of exposition, we focus on the case with only one treated unit, $j = 1$, while the rest,  $j=2,\dots, J+1$, are in the control group. The treated unit receives the treatment or policy of interest at period $T_0+1$, while the units in the control group do not receive any treatment throughout the entire period.\footnote{We focus on this assumption because it is commonly adopted and generally considered to hold in the empirical settings we examine. While some recent studies have sought to relax the no-spillover assumption—for instance, the inclusive Synthetic Control Method proposed by \citet{distefano2024inclusive}, which addresses potential biases arising from spillovers or multiple treated units—our contribution targets a distinct challenge of the traditional SC framework: constructing a robust weighting scheme under the assumption of a valid donor pool.}

Using the potential outcome framework developed in \cite{rubin_1974}, let $Y_{jt}(1)$ and $Y_{jt}(0)$ denote the potential outcome of unit $j$ in period $t$ with and without treatment, respectively. The dynamic treatment effect of interest on the treated unit in period $t$ is given by
\begin{equation}
\delta_{1t} = Y_{1t}(1) - Y_{1t}(0), \quad t = T_0 + 1, \dots, T. \label{eq:dTE}
\end{equation}
We assume $\delta_{1t}$ follows a stationary process, then the average treatment effect (ATE) on the treated is given by
\begin{equation}
\delta_1 = \EE(\delta_{1t}). \quad t \geq T_{0}+1\label{eq:ATE}
\end{equation}
Here, the expectation is taken with respect to the stationary distribution of the dynamic treatment effects, so $\delta_1$ represents the long-run average causal effect on the treated unit across the post-treatment periods.

The fundamental challenge in causal inference when estimating a treatment effect is that we cannot observe both $Y_{jt}(1)$ and $Y_{jt}(0)$ at the same time. Instead, for the treated unit, we can only observe the potential outcome without treatment during the first $T_0$ periods (pre-treatment periods), whereas we only observe the potential outcome with treatment from $T_0+1$ onward. Throughout the entire time, we can only observe the potential outcome without treatment for the control units. The \emph{observed} outcome, $Y_{jt}$ is given by
\begin{align*}
\text{Treated Unit: 
 }    Y_{1t} &= Y_{1t}(0), \quad t = 1, \dots, T_0, \\
    Y_{1t} &= Y_{1t}(1), \quad t = T_0 + 1, \dots, T,\\
\text{Control Units: 
 }        Y_{jt} &= Y_{jt}(0), \quad t = 1, \dots, T, \quad 2 \leq j \leq J+1.
\end{align*}

To estimate the treatment effects for the treated unit, we need to construct the counterfactual outcome, the potential outcome without treatment, $Y_{1t}(0)$.
As noted in \cite{doudchenko_balancing_2016}, many of the estimators in the literature share the following structure for the imputation of $Y_{1t}(0)$:
\begin{equation}\label{eq:cfm}
    Y_{1t}(0) = \mu^0 + \sum_{j=2}^{J+1}\omega_j^0 Y_{jt} + u_{1t} = \mu^0 + Y_t \omega^0 + u_{1t}, \quad t = 1, \dots,T,
\end{equation}
where $\mu^0$ is the intercept highlighting the level difference between the treated and control units, $Y_t = (Y_{2t}, \dots, Y_{J+1, t})$ is a $1\times J$ vector containing all control units' outcomes at time $t$, $\omega^0$ is a $J\times 1$ vector denoting the unknown true weights of $Y_t$, and $u_{1t}$ is a zero mean, finite variance idiosyncratic error term.  As noted in \cite{shen2022heterogeneous}, this model in (\ref{eq:cfm}) is flexible, well approximating various data-generating processes, including linear models, latent factor models, and autoregressive models, and firmly rooted in numerous practical applications. The key idea is that ``as the time series trends are typically driven by a few common factors, a weighted average of control units often provides a good approximation for the counterfactual outcome of the treated unit as if it were under control.''

Note that following \cite{doudchenko_balancing_2016}, (\ref{eq:cfm}) is an extended version of the traditional SC assumption, as this specification further allows for a permanent level difference between the treatment and the (synthetic) control groups, $\mu^0$, which aligns with the traditional DID strategy. Such extension ensures that ``the goal of \(\omega\) is no longer to perfectly match the treatment group, but only to mimic the trend of the treatment group. Even if we don't get a perfect match, we can correct it later with a unit fixed effect (an aspect of DID).''

\subsection{Penalized Regression Methods} \label{subsec:prm}

Estimation of the treatment effects in (\ref{eq:dTE}) and (\ref{eq:ATE}) hinges on the construction of the counterfactual outcome in (\ref{eq:cfm}), which in turn depends on the estimation of $\mu^0$ and $\omega^0_j$. 
Following the literature (e.g., \cite{doudchenko_balancing_2016}), our paper focuses primarily on penalized regression for the estimation of $\mu^0$ and $\omega^0_j$ using the pre-treatment period data by solving the following problem:\footnote{This is also called the vertical regression approach (\cite{athey2021matrix}.)} 
\begin{equation}\label{eq:prob_pen}
    (\widehat{\mu}, \sol) = \argmin_{(\mu, \omega) \in (\RR, \RR^J)} \left\{\frac{1}{2}\sum_{t=1}^{T_0} \left( Y_{1t}-\mu-\sum_{j=2}^{J+1}\omega_j Y_{jt} \right)^2 + \cP(\omega)\right\},
\end{equation}
where $\cP(\omega)$ is a penalty function that penalizes deviations away from zero. The literature has adopted a few variants of $\cP(\omega)$. For example, $\cP(\omega) = \lambda \norm{\omega}_1$ represents Lasso regression, $\cP(\omega) = \lambda \norm{\omega}_2$ represents ridge regression, and $\cP(\omega) = \lambda \left(\alpha \norm{\omega}_1 + \frac{(1-\alpha)}{2} \norm{\omega}_{2}^2\right)$ introduces elastic net, where $\|\cdot\|_1$ and $\|\cdot\|_{2}$ represent $L_1$ and $L_{2}$ norms, respectively.
These methods address challenges such as multicollinearity and overfitting. 

In this paper, we introduce two penalization methods based on the $L_{\infty}$ norm as follows: 
\begin{equation}\label{eq:pen}
    \begin{aligned}
        \cP(\omega) & = \lambda \norm{\omega}_{\infty}, \\
        \cP(\omega) & = \lambda (\alpha \norm{\omega}_1 + (1-\alpha) \norm{\omega}_{\infty}). \\
    \end{aligned}
\end{equation}
where $\|\omega\|_{\infty}$ denotes the maximum absolute value among the components of $\omega$.
Our approach not only shares similar advantageous properties as the conventional ones but also mitigates the risk of overemphasizing a few control units. The penalty function, $\cP(\omega)=\lambda \norm{\omega}_{\infty}$, introduces a \emph{dense} weighting scheme that reduces undue weight concentration and incorporates a broader set of control units are incorporated in the estimation. On the other hand, the penalty function $\cP(\omega)=\lambda (\alpha \norm{\omega}_1 + (1-\alpha) \norm{\omega}_{\infty})$ is an eleastic-net like approach, thereby balancing denseness and sparsity while introducing a \emph{denser} weighting scheme than the conventional SC approaches.

Having obtained the estimates of $\mu^0$ and $\omega^0$, we can construct the dynamic treatment effects, $\delta_{1t}$, as follows:
\begin{equation}\label{eq:delta_t_est}
    \widehat{\delta}_{1t} = Y_{1t} - \widehat{Y}_{1t}(0) = Y_{1t} - \left(\widehat{\mu} + \sum_{j=2}^{J+1}\sol_jY_{jt} \right)
\end{equation}
for $t = T_0 + 1, \ldots, T$. Furthermore, the estimator of ATE on the treated is given by
\begin{equation}\label{eq:delta_est}
    \widehat{\delta}_1 = \frac{1}{T_1}\sum_{t=T_0+1}^{T}\widehat{\delta}_{1t}.
\end{equation}

\noindent where $T_1$ is the number of post-treatment periods.

\subsection{Interpretation of the Constrained Estimator}\label{subsec:eqp}
The introduction of the infinity norm in this context is novel, and its properties warrant further exploration. To ease interpretation, it is beneficial to examine the relationship between a constrained estimator and the ordinary least squares estimator. This exercise will also aid in understanding the geometric properties and theoretical results presented in later sections.

For ease of exposition, we assume that \(Y_{jt}\)'s are centered for each \(j\), so $\mu$ can be omitted without loss of generality. If one is interested in the properties of $\mu$, the results remain unchanged by replacing $Y$ and \(\omega\) with $\tilde{Y} = [\one, Y]$ and $\tilde{\omega} = (\mu, \omega^{\top})^{\top}$. 
Here, we rewrite the least square part of (\ref{eq:prob_pen}) in matrix form:
\begin{equation}\label{eq:prob_pen_wo_mu}
\begin{split}
    \sol & = \argmin_{\omega \in \RR^J} \frac{1}{2}\sum_{t=1}^{T_0} \left( Y_{1t}-Y_t \omega \right)^2 + \cP(\omega)\\
    & = \argmin_{\omega \in \RR^J}\frac{1}{2}(y - Y\omega)^{\top}(y - Y\omega) + \cP(\omega),
\end{split}
\end{equation}
where $y = (Y_{11}, \dots, Y_{1T_0})^{\top}$, and 
$$
Y = 
\begin{pmatrix}
Y_1    \\
Y_2    \\
\vdots \\
Y_{T_0}\\
\end{pmatrix}
=
\begin{pmatrix}
Y_{21} & Y_{31} & \cdots & Y_{J+1,1} \\
Y_{22} & Y_{32} & \cdots & Y_{J+1,2} \\
\vdots   & \vdots   & \ddots & \vdots   \\
Y_{2T_0} & Y_{3T_0} & \cdots & Y_{J+1,T_0} \\
\end{pmatrix}_{T_0 \times J}.
$$ 
Note that the least square estimator \(\lse\) is a special case when $\cP(\omega)=0$, formally defined by:
\begin{equation*}
    \lse = \argmin_{\omega \in \RR^J} \left\{ \frac{1}{2}(y - Y\omega)^{\top}(y - Y\omega) \right\} = \left(Y^{\top}Y\right)^{-1}Y^{\top}y,
\end{equation*}
where \(Y\) is assumed to be of full column rank.

To demonstrate the relationship between \(\sol\) and the least squares estimator, \(\lse\), we first introduce the concept of \textbf{proximal operator}. Let \(f: \RR^J \to \RR \cup \{+\infty\}\) be a closed convex function. 
The operator \(\mathbf{prox}_{\lambda f}: \RR^J \to \RR^J\) associated with scaled function \(\lambda f\) and $\lse$ is defined by 
\begin{equation}\label{eq:prox_op}
    \mathbf{prox}_{\lambda f}(\lse) = \argmin_{x}\left(f(x) + \frac{1}{2\lambda}\|x - \lse\|_2^2\right).
\end{equation}
This definition indicates that \(\mathbf{prox}_{\lambda f}(\lse)\) is a point that balances minimizing \(\lambda f\) and staying close to \(\lse\), while the parameter \(\lambda > 0\) controls the trade-off between the two goals. 

To remove the influence of predictor-induced distortions or transformations, we assume that $Y$ is orthonormal. This assumption allows us to focus solely on the structure imposed by the penalty term. Then, we can show that (\ref{eq:prob_pen_wo_mu}) is equivalent to (\ref{eq:prox_op}) with \(\lambda f = \cP\) provided that \(\cP\) is a convex penalization function. 
Proposition \ref{prop:decomp} shows that \(\sol\) can be decomposed into two components with $\lse$ and its proximal operator. 

\begin{proposition}\label{prop:decomp}
Suppose that \(Y\) is orthonormal and \(\cP(\omega)\) is a convex penalization function, we have
\[
\sol = \lse - \mathbf{prox}_{\cP^*}(\lse),
\]
where \(\cP^*(y) = \sup_{x}(y^{\top}x - \cP(x))\) is the convex conjugate of \(\cP\).

    \noindent \textbf{Case 1:} when \(\cP(\omega) = \lambda \norm{\omega}_{\infty}\), we have
\[
\sol = \lse - \Pi_{\Omega_1}(\lse),
\]
where \(\Omega_1 = \{\omega \mid \|\omega\|_1 \leq \lambda\}\) and \(\Pi_{\Omega_1}(\omega) = \argmin_{\omega \in \Omega_1} \|\omega - \lse\|_2^2\).

    \noindent \textbf{Case 2:} when \(\cP(\omega) = \lambda (\alpha \norm{\omega}_1 + (1-\alpha) \norm{\omega}_{\infty})\), we have
\[
\sol = \lse - \Pi_{\Omega_2}(\lse),
\]
where \(\Omega_2 = \{\omega\mid\sum_{j=1}^J \max(|\omega_{j+1}| - \lambda(1-\alpha), 0) \leq \lambda\alpha\}\) and \(\Pi_{\Omega_2}(\omega) = \argmin_{\omega \in \Omega_2} \|\omega - \lse\|_2^2\).

\end{proposition}

Proposition~\ref{prop:decomp} establishes the connection between \(\sol\) and \(\lse\). While the analytical solution for (\ref{eq:prob_pen_wo_mu}) is not available, we can decompose \(\lse\) into two mutually orthogonal spaces. In particular, we can determine the solution for the problem with \(\cP(\omega)=\lambda\norm{\omega}_{\infty}\) by projecting \(\lse\) onto \(\Omega_1\) and subtracting the projection from \(\lse\). Similarly, we can obtain the solution for the problem with \(\cP(\omega)=\lambda (\alpha \norm{\omega}_1 + (1-\alpha) \norm{\omega}_{\infty})\) by computing \(\lse - \Pi_{\Omega_2}(\lse)\). Below, we use these relationships to illustrate the unique properties of the constrained estimators.

\begin{figure}[htpb]
    \centering
    \includegraphics[width=0.5\textwidth]{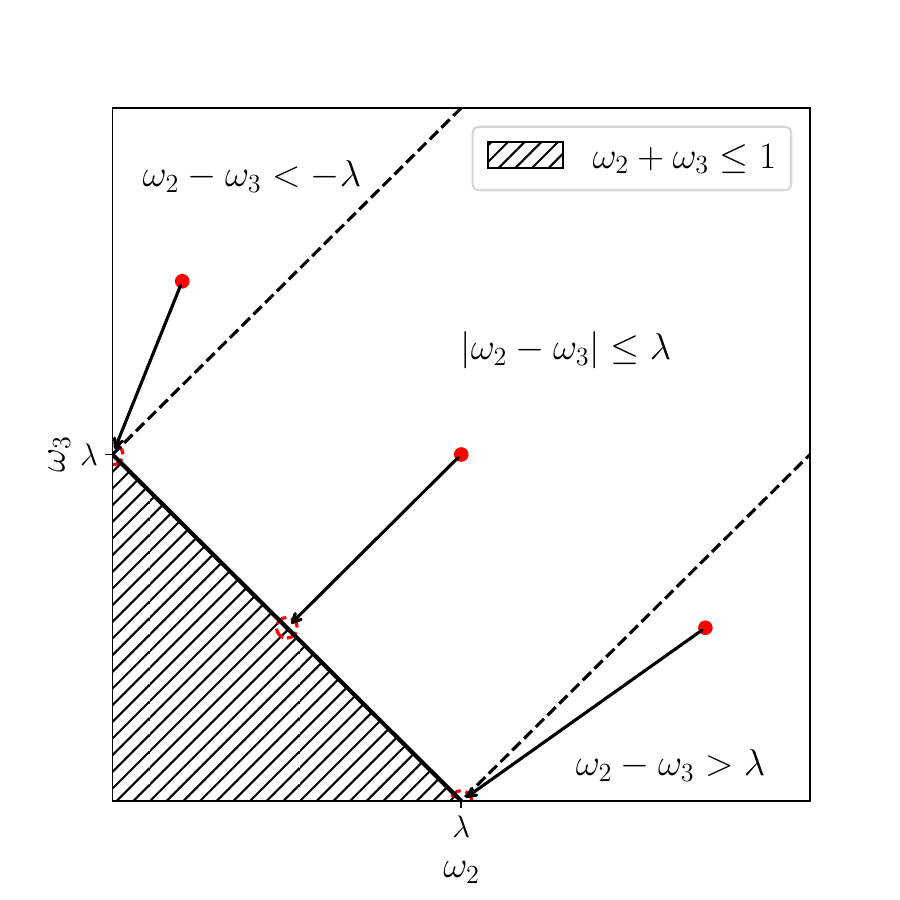}

    \caption{\small Projection of \(\lse\) onto the $L_1$ ball \(\Omega_1\) varies in terms of the relationship between \(\lse_2\) and \(\lse_3\).}
    \label{fig:ex1_pre}
\end{figure}

\noindent \textbf{Illustrative Example 1 ($\cP(\omega)= L_{\infty}$):} Now,  we consider an example where there are two control units, i.e., $J = 2$. Let \(\sol^{\infty}\) denote the minimizer of (\ref{eq:prob_pen_wo_mu}) with $L_{\infty}$ penalty and \(\lse_{j+1}\) denote the weight for the \(j\)-th control unit, i.e., the $j$-th component of \(\lse\). Without loss of generality, we assume that \(\lse_{j+1} > 0\). Figure~\ref{fig:ex1_pre} illustrates the projection onto \(\Omega_1\), and Table \ref{tab:example1} details the projection and solution formulae.

\begin{table}[htpb]
\caption{Analytical Solutions for the case of $J=2$}
\label{tab:example1}
\centering
\begin{tabular}{l|l|l}
\toprule
                       Condition &                             $\Pi_{\Omega_1}(\lse)$ &                                    $\sol^{\infty}$ \\
\midrule
     $\lse_2 - \lse_3 > \lambda$ &                              $(\lambda, 0)^{\top}$ &                       $(\lse_2 - \lambda, \lse_3)$ \\
    $\lse_2 - \lse_3 < -\lambda$ &                              $(0, \lambda)^{\top}$ &                       $(\lse_2, \lse_3 - \lambda)$ \\
$|\lse_2 - \lse_3| \leq \lambda$ & $\left(\frac{(\lse_2 - \lse_3 + \lambda)}{2}, \frac{(\lse_3 - \lse_2 + \lambda)}{2} \right)$ & $\left(\frac{\lse_2 + \lse_3 - \lambda}{2},  \frac{\lse_2 + \lse_3 - \lambda}{2} \right)$\\
\bottomrule
\end{tabular}
\end{table}

\begin{table}[htpb]
\caption{Estimation formulae for Lasso, ridge, and $L_{\infty}$. \(\sign(x)\) denotes the sign of \(x\) (\(\pm 1\)), \(x_+\) denotes the ``positive'' part of \(x\).}
\label{tab:sol}
\centering
\begin{tabular}{cc}
\toprule
Method & Formula ($j = 2$) \\
\midrule
Lasso & $\sign(\lse_2)(|\lse_2| - \lambda)_{+}$ \\
Ridge & $\lse_2 / (1 + 2\lambda)$ \\
$L_\infty$ & $\sign(\lse_2) \left(|\lse_2|- \frac{1}{2}\left[\min(\max(|\lse_2| - |\lse_3|, -\lambda), \lambda) + \lambda\right]\right)_{+}$ \\
\bottomrule
\end{tabular}
\end{table}

\begin{figure}[ht]
    \centering
    \includegraphics[width=.9\textwidth]{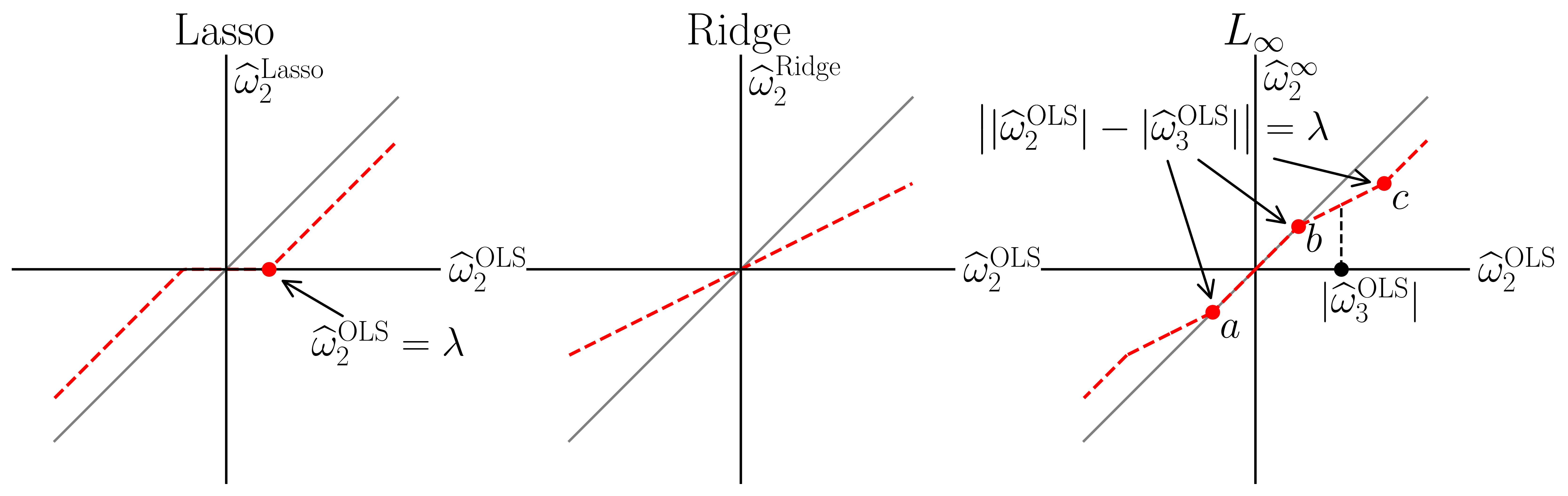}
    \caption{\small The dashed red line represents estimators of $\omega_2$ given by Lasso, ridge, or $L_{\infty}$ method. The 45$^{\circ}$ line in gray represents the estimator without penalty for reference.}
    \label{fig:ex1}
\end{figure}

In Table \ref{tab:sol}, we further inspect the  \(\sol^{\infty}\) solution and provide comparisons to the solutions obtained from Lasso and ridge regression methods. Figure~\ref{fig:ex1} visualizes such comparisons among the constrained estimators \(\sol_2^{\text{Lasso}}\), \(\sol_2^{\text{Ridge}}\), and \(\sol_2^{\infty}\), and how $\lambda$ regularizes the magnitudes of these estimators. The \textbf{Lasso} regression (left panel) reduces \(|\lse_2|\) by $\lambda$, and the highlighted point marks the location where $\lse_2 = \lambda$, at which $\sol_2^{\text{Lasso}}$ becomes zero, hence sparsity in a higher dimensional setting. 
The \textbf{Ridge} regression (middle panel), on the other hand, uniformly shrinks the estimates by a constant factor of $(1 + 2\lambda)^{-1}$. 

Our proposed estimator based on $L_{\infty}$ (right panel) demonstrates a more complex penalization behavior, which is a function of the difference between \(|\lse_2|\) and \(|\lse_3|\). Moreover, an important, unique feature here is that the penalty for one weight also depends on the magnitudes of other weights. To see this, suppose that \(|\lse_3|\) is fixed. Points \(a\) and \(b\) indicate the positions where \(|\lse_3| - |\lse_2| = \lambda\), and point \(c\) marks where \(|\lse_2| - |\lse_3| = \lambda\).
To the right of point \(c\), $\lse_2$ is penalized because $|\lse_2| - |\lse_3| > \lambda$ and \(\sol_3\) is unpenalized; in other words, the maximum weight with the excessive distance to other weights exceeding $\lambda$ will be penalized, but not the others. Symmetrically, between points $a$ and $b$, $\lse_2$ is unpenalized, while \(\lse_3\) is penalized as $|\lse_3| - |\lse_2| > \lambda$. Finally, in the region between the points $b$ and $c$, both $\lse_2$ and $\lse_3$ are penalized because the difference in their absolute values is less than $\lambda$.  Readers may verify that the above process is fully consistent with the last row of Table~\ref{tab:sol}.

\begin{figure}[htb!]
    \centering
    \includegraphics[width=.8\textwidth]{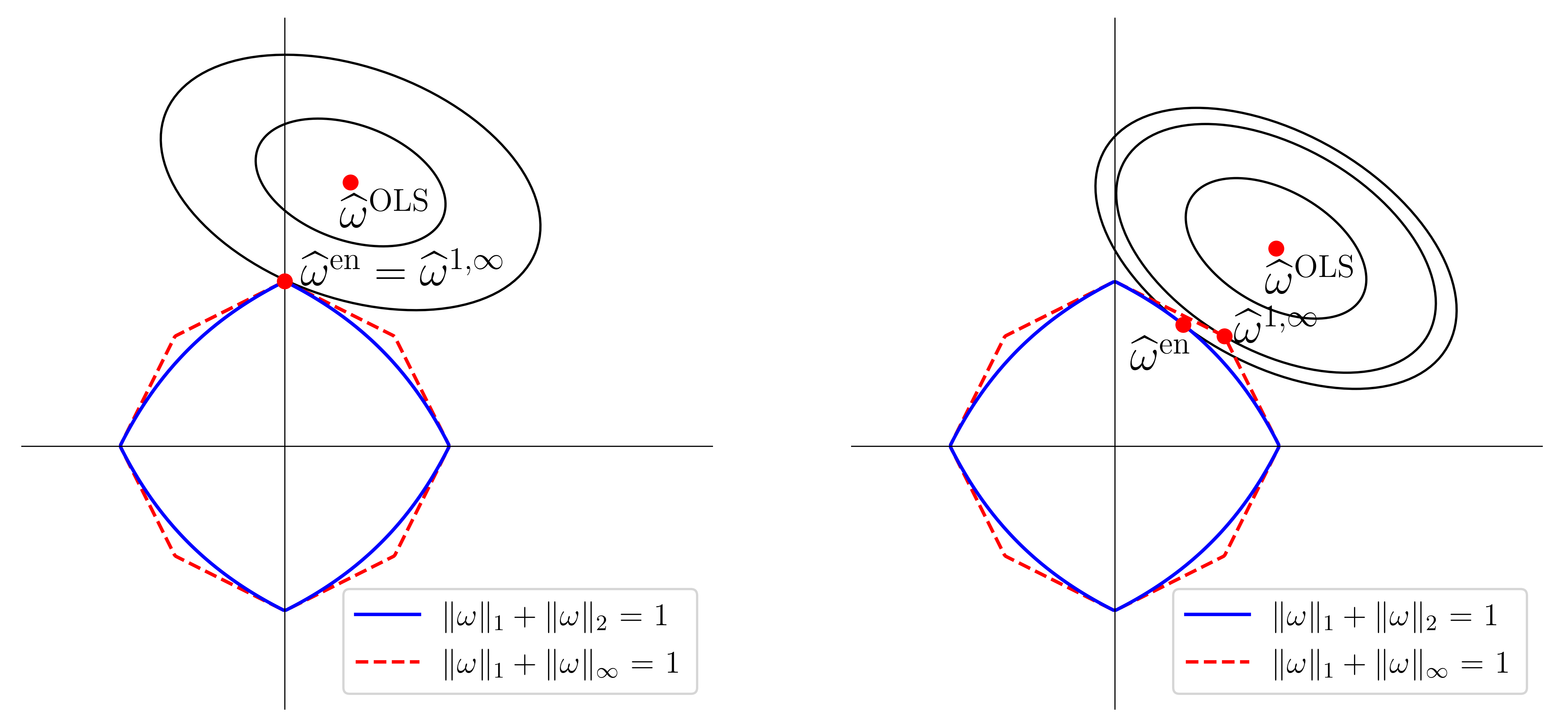}
    \caption{\small 
     The solid blue line indicates the points where $\|\omega\|_1 + \|\omega\|_2 = 1$, while the dashed red line represents the points where $\|\omega\|_1 + \|\omega\|_{\infty} = 1$.    
    \textbf{Left Panel}: when the loss function is tangent to both constraint lines at the same point, elastic net and $L_1 + L_{\infty}$ yield the same sparse solution, i.e., $\sol^{\textrm{en}} = \sol^{1, \infty}$;
    \textbf{Right Panel}: when the loss function is tangent to the constraint lines at different points, elastic net and $L_1 + L_{\infty}$ produce different solutions, i.e., $\sol^{\textrm{en}} \neq \sol^{1, \infty}$.}
    \label{fig:ex2}
\end{figure}

\noindent \textbf{Illustrative Example 2 \(\cP(\omega)=\lambda (\alpha \norm{\omega}_1 + (1-\alpha) \norm{\omega}_{\infty})\):}  Next, we consider (\ref{eq:prob_pen_wo_mu}) with the composite penalty function \(\cP(\omega)=\lambda (\alpha \norm{\omega}_1 + (1-\alpha) \norm{\omega}_{\infty})\). 
Since the projection space \(\Omega_2\) is complex, the analytical solution \(\sol\) for this function is not available. 
However, based on the Lagrangian multiplier, we know that the original problem can be converted into a constrained optimization problem:
\begin{equation}\label{eq:prob_l1_inf}
\begin{split}
\min \quad & \frac{1}{2}(y - Y\omega)^{\top}(y - Y\omega) \\
\text{s.t.:}\quad & \alpha\|\omega\|_1 + \|\omega\|_{\infty} \leq c,
\end{split}
\end{equation}
where \(c\) is a constant uniquely determined by \(\alpha\) and \(\lambda\) together.
Let \(\sol^{1,\infty}\) and \(\sol^{\textrm{en}}\) denote the minimizers of (\ref{eq:prob_pen_wo_mu}) with \(L_1 + L_{\infty}\) and elastic net \citep{zou_regularization_2005}\footnote{The elastic net combines the strengths of Lasso and ridge regression by incorporating both $L_1$ and $L_2$ penalties.}, respectively.
Again, we assume that \(J=2\) and provide an example in Figure~\ref{fig:ex2} to demonstrate a case where elastic net and \(L_1 + L_{\infty}\) method yield the same result, as well as a case where they differ.

The left panel of Figure~\ref{fig:ex2} demonstrates that both elastic net and $L_1 + L_{\infty}$ method can lead to the same sparse solution when the loss contour is tangent to the constraint boundaries at the same location. In addition, the fact that \(\sol^{1, \infty}_2 = 0\) indicates that $L_1 + L_{\infty}$ is capable of producing sparse solutions when such solutions are optimal. 
The right panel highlights a key difference between elastic net and $L_1 + L_{\infty}$ method. 
In this case, elastic net produces a solution $\sol^{\textrm{en}}$ whose components have unequal absolute values, while $L_1 + L_{\infty}$ produces a different solution $\sol^{1, \infty}$ whose components have equal absolute values. 
This distinction suggests that using $L_1 + L_{\infty}$ to solve (\ref{eq:prob_pen_wo_mu}) is advantageous when the goal is to assign more balanced weights to the control units. 
Both elastic net and \(L_1 + L_{\infty}\) constrain the largest absolute value of \(\sol\). However, while elastic net primarily promotes sparsity, \(L_1 + L_{\infty}\) encourages both sparsity and more uniform magnitudes among nonzero coefficients.

Our illustrative examples above demonstrate that, even with \(J=2\) and an orthonormal design matrix \(Y\), the solution for (\ref{eq:prob_pen_wo_mu}) with \(L_{\infty}\) or \(L_1 + L_{\infty}\) penalty remains complex. In fact, since closed-form solutions for the problems with \(L_{\infty}\) and \(L_1 + L_{\infty}\) penalties are generally not available, we must employ a more suitable numerical algorithm to solve it, rather than the ones used for Lasso regression or ridge regressions.

\section{Implementation}\label{sec:imp}

A few convex optimization algorithms are available and commonly used, including gradient-based methods, coordinate descent methods, and Interior Point Methods (IPMs) to solve (\ref{eq:prob_pen}). Here, we adopt the IPM for the following reasons. First, gradient-based methods are suitable for smooth convex functions but are less effective when the \(L_1\) and \(L_{\infty}\) norms introduce non-differentiable points in the search space. Second, traditional coordinate descent methods iteratively update each coordinate while holding the others fixed. However, as shown in Figure~\ref{fig:ex1}, the penalization induced by \(L_{\infty}\) is highly complicated even when \(J=2\), and more importantly, it depends on \emph{all} weights (coordinates). The updating procedures become computationally burdensome, if not infeasible, when \(J\) is relatively large. 
In light of the numerous constraints and the complex feasible region in our case, IPMs emerge as the most suitable choice, as they are particularly effective for such challenges \citep{wright1997primal}. 
In this section, we reformulate (\ref{eq:prob_pen}) into the format used by IPMs and introduce its basic concept and steps.

\subsection{Reformulation of the Convex Optimization Problem}

The standard formulation of a convex optimization problem used by IPMs is given by: 
\begin{equation*}
\begin{split}
\min \quad & (1/2)x^{\top} Q x + q^{\top} x \\
\text{s.t.:} \quad & G_{m\times (J+1)}x \leq h_{m \times 1},
\end{split}
\end{equation*}
where $x \in \RR^{J+1}$ is the variable, $Q \in \RR^{(J+1)\times (J+1)}$ represents the quadratic coefficient matrix, $q \in \RR^{J+1}$ denotes the linear coefficients, and $G x \leq h$ represents \(m\) inequality constraints. 
Let $c$ be an upper bound of \(\|\omega\|_{\infty}\). 

In our context, (\ref{eq:prob_pen}) with the \(L_{\infty}\) penalty is equivalent to
\begin{equation}\label{eq:prob_inf2}
\begin{split}
\min \quad & \frac{1}{2}(y - \tilde{Y}\tilde{\omega})^{\top}(y - \tilde{Y}\tilde{\omega}) + \lambda c \\
\text{s.t.:} \quad & -c\one \leq \omega \leq c\one,
\end{split}
\end{equation}
where $\tilde{Y} = [\one, Y]$ and $\tilde{\omega} = (\mu, \omega^{\top})^{\top}$. Similarly, let $d = (d_1, \dots, d_{J})^{\top}$ be an upper bound vector for $|\omega|$. 
Thus, (\ref{eq:prob_pen}) with the \(L_1 + L_{\infty}\) penalty is equivalent to 
\begin{equation}\label{eq:prob_l1_inf2}
\begin{split}
\min \quad & \frac{1}{2}(y - \tilde{Y}\tilde{\omega})^{\top}(y - \tilde{Y}\tilde{\omega}) + \lambda \left(\alpha \sum_{j=1}^{J}d_j + (1-\alpha)c\right) \\
\text{s.t.:} \quad & -c\one \leq \omega \leq c\one, -d \leq \omega \leq d.
\end{split}
\end{equation}
IPMs are particularly effective for convex optimization, especially in problems with numerous constraints and complex feasible regions.

\subsection{Interior Point Methods}\label{subsec:ipm}

For a general convex optimization problem, IPMs break the matrix constraints \(Gx \leq h\) into \(m\) inequalities:
\begin{equation*}
\begin{split}
\min \quad & (1/2)x^{\top} Q x + q^{\top} x \\
\text{s.t.:} \quad & g_ix - h_i \leq 0, \quad i = 1, \ldots, m.
\end{split}
\end{equation*}
where $g_i$ is the $i$-th row of $G$ and $h_i$ is the $i$-th component of $h$.
IPMs approximate the original problem by minimizing the function \(\phi(x, \gamma)\), defined as
$$
\phi(x, \gamma) = \gamma \left((1/2)x^{\top} Q x + q^{\top} x\right) - \sum_{i=1}^{m} \log(-g_i x + h_i), 
$$
where \(\gamma\) is a positive parameter. 
The logarithmic term ensures the positivity of \(-g_i x + h_i\), which guarantees that the original inequality constraint \(g_i x - h_i \leq 0\) is satisfied. 
To find the minimizer of $\phi(x, \gamma)$, IPMs use Newton's method when \(Q\) is positive definite. When \(Q\) is not positive definite, IPMs adopt a first-order method instead, such as gradient descent, which updates the point by moving in the direction opposite to the gradient of the function at the current point.
The method iteratively solves a sequence of barrier problems for an increasing sequence of parameters \(\{\gamma_k\}_{k=1}^{\infty}\). 
In iteration \(k\), the minimizer of \(\phi(x, \gamma_k)\), denoted as \(x_k\), is determined by using \(x_{k-1}\) from iteration \(k-1\) as the starting point.
As $\gamma_k \to \infty$, \(\min_x\phi_{x, \gamma}\) approaches the solution of (\ref{eq:prob_pen}) arbitrarily close.
Thus, the minimizer of \(\phi(x, \gamma)\) traces a path that converges to the solution of (\ref{eq:prob_pen}). 
The algorithm terminates when a predefined convergence criterion is met. 
For our numerical analysis, we employ the \texttt{CVXOPT} library embedded in Python, which implements IPMs using the central path following approaches \citep{andersen2003implementing, tutuncu2003solving} to implement IPMs.

\begin{algorithm}[hb]
\caption{Estimation of Dynamic Treatment Effects and ATE using $L_{\infty}$ or $L_1 + L_{\infty}$}
\label{alg:est_ate}
\begin{algorithmic}[1]
\REQUIRE Observations $\{ Y_{jt} \}_{j=1,\dots,J+1; \, t=1,\dots,T}$, with $y \in \RR^{T_0 \times 1}$ as pre-treatment data for the treatment group and $Y \in \RR^{T_0 \times J}$ as pre-treatment data for the control group. Ensure that $Y^{\top}Y$ is invertible.
\STATE Generate a sequence of hyper-parameters: $\{(\lambda_i, \alpha_i)\}_{i=1}^{n}$ \COMMENT{$\{\lambda_i\}_{i=1}^n$ for $L_{\infty}$}
\FOR{$i=1$ to $n$}
    \STATE Randomly split $\{1,\dots,T_0\}$ into $K$ folds \COMMENT{$K$ is up to the user's choice.}
    \FOR{each fold $k = 1,\dots,K$}
        \STATE Let $Y_{(-k)}$ and $y_{(-k)}$ denote the data excluding $k$-th fold, i.e., $Y$ and $y$ excluding rows in the $k$-th fold
        \STATE Solve (\ref{eq:prob_inf2}) or (\ref{eq:prob_l1_inf2}) using $y_{(-k)}$ and $Y_{(-k)}$ to obtain $(\widehat{\mu}^{(i,k)}, \sol^{(i,k)})$ 
        \STATE Compute $\widehat{\delta}_{1t}^{(i)}$ for $t \notin k$-th fold
    \ENDFOR
    \STATE Compute the RMSE $\sqrt{\frac{1}{T_0}\sum_{t=1}^{T_0} \left(\widehat{\delta}_{1t}^{(i)}\right)^2}$ for iteration $i$. \COMMENT{$\delta_{1t} = 0$, \; $t=1,\dots,T_0$}
\ENDFOR
\STATE Select $(\lambda^*, \alpha^*)$ that minimizes RMSE, and solve (\ref{eq:prob_inf2}) or (\ref{eq:prob_l1_inf2}) with $(\lambda^*, \alpha^*)$, $y$, and $Y$ to obtain $(\widehat{\mu}, \sol)$
\STATE Calculate dynamic treatment effect $\widehat{\delta}_{1t}$ for $t = T_0+1, \dots, T$ as in (\ref{eq:delta_t_est}) and obtain average treatment effect estimate $\widehat{\delta}_1$ as in (\ref{eq:delta_est})
\end{algorithmic}
\end{algorithm}

\subsection{Choosing the tuning parameter}

Selecting appropriate values for $\lambda$ and $\alpha$ is essential to estimate \(\omega\). First, we determine their search range. The parameter $\alpha$, which controls the trade-off between $L_1$ norm and \(L_{\infty}\) norm, naturally ranges between 0 and 1. We generate an equally spaced sequence within this range. 
For $\lambda$, we adopt the approach outlined in the elastic net \texttt{R} package \citep{friedman2010glmnet}. Specifically, we calculate the inner products between the standardized \(Y_{jt}, j = 2,\dots, J+1, t = 1,\dots, T_0\) and \(Y_{1t}, t=1,\dots,T_0\), taking the largest value and dividing it by \(T_0\) to obtain the maximum \(\lambda\), denoted by $\lambda_{\textrm{max}}$. Subsequently, we compute the minimum $\lambda_{\textrm{min}}$ as $\lambda_{\textrm{min}} = \lambda_{\textrm{max}} * \epsilon$, where \(\epsilon\) is typically set as \(0.0001\). Finally, we generate a sequence of $\lambda$ values equally spaced on the logarithmic scale of \([\lambda_{\textrm{min}}, \lambda_{\textrm{max}}]\).
For each pair of \((\lambda, \alpha)\) in \([\lambda_{\textrm{min}}, \lambda_{\textrm{max}}] \times [0, 1]\), we estimate the treatment effect $\delta_j$ for \(j = 2,\dots,J+1\). 
To achieve this, we employ a \(K\)-fold cross-validation strategy. The control units are divided into \(K\) folds, with each fold alternatively treated as the experimental group, while the remaining folds serve as the control group. Let $Y_{(-k)}$ and $y_{(-k)}$ denote the data excluding the $k$-th fold, i.e., $Y$ and $y$ with rows from the $k$-th fold removed.
We then estimate the intercept, weights, and treatment effects \(\widehat{\delta}_j(\alpha, \lambda)\) using $Y_{(-k)}$ and $y_{(-k)}$. The root mean squared error (RMSE) across all control units is computed using the formula:
\[
\textrm{RMSE}(\alpha, \lambda) = \sqrt{\frac{\sum_{j=2}^{J+1}\widehat{\delta}_j^2(\alpha, \lambda)}{J}}
\]
This process is repeated for every pair of $\lambda$ and $\alpha$. The last step is to select the pair that minimizes the RMSE.

\section{Theoretical Results}\label{sec:theory}

As noted in \cite{li_statistical_2020}, despite the widespread application of the SC method, there has been limited formal inference theory for the SC estimator of the average treatment effect (ATE) in long panel settings, i.e., when both the pre- and post-treatment periods are large and the underlying conditions are general. Most existing inference procedures rely on placebo tests or permutation-based methods, which are typically calibrated for short post-treatment horizons and do not offer valid asymptotics in longer panels. This remains an active area of research. Our approach contributes to this emerging literature, complementing several other recent methods that share similar goals.\footnote{For example, recent literature has proposed powerful, distribution-free methods. \cite{lei2021conformal} develop a framework based on conformal inference to construct reliable prediction intervals for counterfactuals with guaranteed finite-sample coverage, offering a robust alternative to traditional asymptotic approaches.} Building on Li (2020), we contribute to this literature by deriving theoretical results that justify inference for the SC ATE estimator under more general and realistic conditions. Specifically, we analyze the theoretical properties of the estimated coefficients \(\sol\) and ATE \(\widehat{\delta}_1\). 
We begin by establishing the distributional results of $\sqrt{T_0}(\sol - \omega^0)$ under weak dependence. 
Next, we add additional regularity conditions to derive the asymptotic distribution of \(\sqrt{T_1}(\widehat{\delta}_1 - \delta_1)\).
As in the previous sections, we consider that $\mu$ is omitted from (\ref{eq:cfm}).

\subsection{Asymptotic Results for the Weights}

We assume that the idiosyncratic errors $\{u_{1t}\}_{t=1}^{T}$ are weakly dependent (\citealp{doukhan1999new}). 
Let $p \in \NN^+$ be a positive integer and denote $\LL^{\infty}$ as the set of measurable bounded functions on $\RR^p$, with its norm written as $\|\cdot\|_{\infty}$. 
Moreover, we endow $\RR^p$ with the norm $\|(x_1, \dots, x_p)\| = \|x_1\| + \cdots + \|x_p\|$. Let $h: \RR^p \to \RR$ be a function defined on $\RR^p$, we set
\begin{equation*}
    \Lip (h) = \sup_{x\neq y} \frac{|h(x) - h(y)|}{\|x-y\|}
\end{equation*}
as the Lipschitz modulus of $h$.
Define
\begin{equation*}
    \cL = \cup_{p=1}^{\infty} \{h\in \LL^{\infty} (\RR^p, \RR); \Lip(h)<\infty, \|h\|_{\infty} \leq 1\}.
\end{equation*} 
where $\cL$ consists of measurable, bounded functions with a finite Lipschitz constant.

\begin{definition}\label{def:weak_dep}
The sequence of random variables $\{u_{1t}\}_{t=1}^{T}$ is called $\mathbf{s}$\textbf{-weakly dependent}, if for some sequence $\{\epsilon_r\}_{r \in \NN}$ decreasing to zero at infinity and any $p+q$-tuples $(t_1, \dots, t_p, s_1, \dots, s_q)$ with $t_1 \leq \dots \leq t_p < t_p + r \leq s_1 \leq \dots \leq s_q$, then $s$-weak dependence conditions are defined for functions $h$ and $k$ defined on $\RR^p$ and $\RR^q$ through the inequality
\begin{equation*}
\left| \Cov(h(u_{1t_1}, \dots, u_{1t_p}), k(u_{1s_1}, \dots, u_{1s_q})) \right| \leq q \Lip(k)\epsilon_r,
\end{equation*}
if $k \in \cL$, $h \in \LL^{\infty}$, and $\|h\|_{\infty} \leq 1$.
\end{definition}

The weak dependence conditions state that the past and the future of a sequence of random variables become irrelevant to each other as they are increasingly far apart (the gap $r$ grows). 
In Supplementary Appendix~B.1, we explore the relationship between weak dependence, stationarity, and strong mixing, and we provide examples of weakly dependent sequences, such as ARMA processes.
Before establishing the distributional property of $\sol$, we introduce two regularity assumptions:

\begin{assumption}\label{as:weak_dep}
    We assume that the error sequence $\{u_{1t}\}_{t=1}^T$ is $s$-weakly dependent as in Definition \ref{def:weak_dep}, and
    \begin{align*}
    & (T_0 M_{T_0} + T_0^{2/3}) M_{T_0}\phi_{T_0} \to 0, \quad
    T_0 M_{T_0} \sum_{t=1}^{T_0} \min(L_{T_0}\epsilon_t, \Phi_{T_0}) \to 0, \\
    & \text{and} \quad T_0 \sum_{t=1}^{T_0} \min(M_{T_0}L_{T_0}\epsilon_t, \Phi_{T_0}) \to 0,
    \end{align*}
    where
    \begin{align*}
    & L_{T_0} = \sup_{1\leq t\leq T_0}|Y_t \omega|/\sqrt{T_0}, \quad M_{T_0} = \sup_{1\leq t\leq T_0}|Y_t \omega u_{1t}|/\sqrt{T_0}, \\
    & \phi_{T_0} = \sup_{1\leq t\leq T_0}\EE[|Y_t \omega u_{1t}|]/\sqrt{T_0}, \quad \text{and} \quad \Phi_{T_0} = \sup_{1 \leq t \neq s \leq T_0} \EE[| \omega^{\top}Y_t^{\top} Y_s \omega u_{1t}u_{1s}|] / T_0.
    \end{align*}
\end{assumption}

Assumption \ref{as:weak_dep} enables the application of the central limit theorem under the $s$-weak dependence condition, which is essential to derive the result of asymptotic normality in Theorem~\ref{thm:w_dist}. We also provide another assumption under the $a$-weak dependence condition in Supplementary Appendix~B.2.

\begin{assumption}\label{as:cov_con}
    Suppose that \(\{Y_t\}_{t=1}^{T}\) is a (weakly) stationary process such that the following hold as \(T \to \infty\):
    \begin{align}
        \frac{1}{T_0} Y^{\top} Y 
        &= \frac{1}{T_0} \sum_{t=1}^{T_0} Y_t^{\top} Y_t \to C_1, \nonumber \\
        \frac{1}{T_0} Y^{\top}\Sigma Y 
        &= \frac{1}{T_0} (\Sigma^{\frac{1}{2}}Y)^{\top}(\Sigma^{\frac{1}{2}} Y) 
        = \frac{1}{T_0} \sum_{t=1}^{T_0} Y_t^{*\top} Y_t^* \to C_2, \nonumber \\
        \frac{1}{T_1} \sum_{t=1}^{T_1} Y_t 
        &\to \mathbb{E}(Y_t). \nonumber
    \end{align}
    \(Y_t^*\) is the \(t\)-th row of the matrix \(\Sigma^{\frac{1}{2}} Y\), and \(C_1\), \(C_2\) are positive definite matrices.
\end{assumption}

Assumption~\ref{as:cov_con} ensures that the sample second moment matrices of the process \(Y_t\) and its transformed version \(\Sigma^{\frac{1}{2}}Y_t\) converge to well-behaved, positive definite matrices \(C_1\) and \(C_2\) as \(T_0\) grows. In other words, the assumption guarantees that the distribution of \(Y_t\) maintains a consistent mean and covariance structure in the long run.
Denote \(u_1 = (u_{11}, \dots, u_{1T_0})^{\top}\), we have \(\lse - \omega^0 = (Y^{\top}Y)^{-1}Y^{\top}u_1\). These two assumptions ensure the convergence of \((Y^{\top}Y)^{-1}Y^{\top}u_1\).

\begin{theorem}[Distributional Property of $\sol$]\label{thm:w_dist}
Let $\sol$ be the estimator obtained by solving the problem in (\ref{eq:prob_pen}) with \(L_1 + L_{\infty}\) and $\omega^0$ be the true coefficient vector. Assume that Assumptions \ref{as:weak_dep} and \ref{as:cov_con} are satisfied. If $\lambda_{T_0}/\sqrt{T_0} \to \lambda_0 \geq 0$, then the estimator $\sol$ has the following property:
\begin{equation}
    \sqrt{T_0}(\sol - \omega^0) \xrightarrow{d} \argmin_{\omega}V(\omega),
\end{equation}
where 
\begin{equation*}
        V(\omega) = -2\omega^{\top}Z + \omega^{\top} C_1 \omega + \lambda_0\left\{\alpha \sum_{j=2}^{J+1} [\omega_j \sign(\omega^0_j)I(\omega^0_j\neq 0) + |\omega_j|I(\omega^0_j = 0)] + (1-\alpha) s^{\top}\omega \right\},
\end{equation*}
$Z$ has a $J$ dimensional $N(\zero, C_2)$ distribution, and $s$ is a $J\times 1$ vector with $j$-th element being $\sign(\omega_{j+1}^0)$ if $\omega_{j+1}^0$ has the largest absolute value and 0 otherwise. In particular, when $\lambda_0 = 0$, we have
\begin{equation*}
    \sqrt{T_0}(\sol - \omega^0) \stackrel{d}{\to} N(\zero, C_1^{-1} C_2 C_1^{-1}).
\end{equation*}
Note: For $\lambda_0$ to be 0, we can choose $\lambda_{T_0} = O\left(\frac{1}{\log T_0 \sqrt{T_0}}\right)$ or any other \(\lambda_{T_0}\) with order faster than $1/\sqrt{T_0}$.
\end{theorem}

The function \(V(\omega)\) consists of two terms. The first term, \(-2\omega^{\top}Z + \omega^{\top} C_1 \omega\), corresponds to ordinary least squares (OLS) estimation. When \(\lambda_{T_0} \to 0\) as \(T_0 \to \infty\), \(\sqrt{T_0}(\sol - \omega^0)\) converges in distribution to \(N(\zero, C_1^{-1} C_2 C_1^{-1})\), which is also the limiting distribution of \(\sqrt{T_0}(\lse - \omega^0)\).
The second term originates from the penalty and influences the behavior of \(\sol\). Specifically, the first component in the second term, \(\sum_{j=2}^{J+1} [\omega_j \sign(\omega^0_j)I(\omega^0_j\neq 0) + |\omega_j|I(\omega^0_j = 0)]\), enforces sparsity by shrinking \(\omega_j\) toward zero when \(\omega^0_j = 0\) and encourages \(\omega_j\) to align with the sign of \(\omega^0_j\) when \(\omega^0_j \neq 0\).
The second component in the second term, \((1-\alpha) s^{\top}\omega\), pushes \(\sol_{j+1}\) toward zero if \(\omega^0_{j+1}\) has the largest absolute value among all \(j\).

\subsection{Asymptotic Results for the Average Treatment Effects}

We establish the asymptotic distribution of \(\sqrt{T_1}(\widehat{\delta}_1 - \delta_1)\) under both stationary conditions and certain nonstationary scenarios, including trend-stationary and unit-root processes, which correspond to our simulation studies and empirical applications. This section primarily addresses the stationary case, and we detail the theoretical developments for the trend-stationary and unit-root nonstationary scenarios in Supplementary Appendices~D.3 and D.4.

\begin{assumption}\label{as:v_norm}
    Let $v_{1t}=\delta_{1t} - \delta_1+u_{1t}$. We assume that $v_{1t}$ has zero mean and satisfies 
    \[T_1^{-1} \sum_{t=T_0+1}^{T} v_{1t} \to N(0, \sigma^2_v),\]
    where 
    \[\sigma^2_v = \lim_{T_1 \to \infty}T_1^{-1}\sum_{t=T_0+1}^{T}\sum_{s=T_0+1}^{T}\EE(v_{1t} v_{1s}).\]
\end{assumption}

\begin{assumption}\label{as:y0_beta}
    Let $\phi=\lim_{T_0, T_1 \to \infty} \sqrt{T_1/T_0}$ be a finite nonnegative constant.
    Let $\bar{Y}_t = (Y_t, \delta_{1t}d_t)^{\top} \in \RR^{J+2}$ for $t=1, \dots, T$, where $d_t = 0$ if $t \leq T_0$ and $d_t = 1$ if $t \geq T_0+1$. 
    Assume that $\{\bar{Y}_t\}_{t=1}^{T_0}$ and $\{\bar{Y}_t\}_{t=T_0+1}^{T}$ are both stationary processes.  
    Define 
    $$\rho(\tau) = \max_{1\leq t\leq T}\max_{1\leq i,j \leq J+2}\left|\Cov(\bar{Y}_{it}, \bar{Y}_{j, t+\tau})/\sqrt{\Var(\bar{Y}_{it})\Var(\bar{Y}_{j, t+\tau})}\right|.$$
    Then there exists positive constants $C>0$ and $0<\eta<1$ such that $\rho(\tau)\leq C\eta^{\tau}$. 
\end{assumption}

Assumption~\ref{as:v_norm} implies that the sum of the errors in the post-treatment period also converges to a normal distribution with variance $\sigma^2_v$.
Some commonly used time series structures, such as AR($p$) and ARMA($p, q$) models, satisfy this assumption. 
Assumption~\ref{as:y0_beta} states that the correlation between any two units \(\bar{Y}_{it}, \bar{Y}_{j, t+\tau}\) decays exponentially as the time gap \(\tau\) increases.
This assumption ensures that the estimator $\sol$ obtained using the pre-treatment data is asymptotically independent of the sample average of the post-treatment errors. In addition, the parameter $\phi$ reflects the relationship between two time periods, $T_1$ and $T_0$. Both can go to infinity, with $T_0$ growing at a faster rate.

\begin{theorem}\label{thm:ate_dist}
    Under Assumptions~\ref{as:weak_dep} -- \ref{as:y0_beta}, we have
    \begin{equation*}
    \sqrt{T_1}(\widehat{\delta}_1 - \delta_1) \stackrel{d}{\to} -\phi \EE(Y_t)\argmin_{\omega}V(\omega) + Z_2,
    \end{equation*}
    where $\phi=\lim_{T_0, T_1 \to \infty}\sqrt{T_1/T_0}$, $Z_2$ is independent with $\argmin_{\omega}V(\omega)$ and distributed as $N(0, \sigma^2_v)$, and $\sigma^2_v = \lim_{T_1 \to \infty}T_1^{-1}\sum_{t=T_0+1}^{T}\sum_{s=T_0+1}^{T} \EE(v_{1t} v_{1s})$.
\end{theorem}
This theorem establishes the limiting distribution of $\sqrt{T_1}(\widehat{\delta}_1 - \delta_1)$, which characterizes how the estimator of $\delta_1$ behaves as the sample size $T_1$ grows large. The distribution consists of two distinct components.
The first component arises from the variation in estimating $\omega^0$. This term is influenced by the pre-treatment data and reflects the uncertainty associated with estimating the parameter $\omega^0$.
The second component captures the variance of the post-treatment errors, which is due to the sample average of $\{u_t\}_{t=T_0+1}^T$.
A key observation is that when $\phi = 0$, meaning that $T_0$ grows faster than $T_1$, the first component vanishes. In other words, with a sufficiently large amount of pre-treatment data, the variation in the estimator of the average treatment effect is primarily driven by the post-treatment observations.

Building on the projection theory developed by  \cite{li_statistical_2020}, we also derive a limiting distribution for \(\sqrt{T_1}(\widehat{\delta}_1 - \delta_1)\) under both trend-stationary and nonstationary cases. The result is characterized by integrals of Brownian motions. For detailed derivations and precise definitions, please refer to Supplementary Appendix~D.3 and D.4.

\section{Simulation}\label{sec:sim}

In this section, we compare our proposed $L_{\infty}$ and $L_1 + L_{\infty}$ methods with several existing approaches, including the conventional SC method and the SC method based on various regularization methods (Lasso, Ridge, and Elastic Net regressions). We consider different data-generating processes (DGPs) for the treated unit, each of which specifies the treated outcome as a weighted combination of the potential outcomes without the treatment: 
$$
Y_{1 t}= \begin{cases}\sum_{j=2}^{J+1} \omega_j Y_{jt}(0)+u_{1t} & \text {if } t \leq T_0, \\ \delta_{1t}+\sum_{j=2}^{J+1} \omega_j Y_{jt}(0)+u_{1t} & \text {if } t>T_0,\end{cases}
$$
where $u_{1t}$ may be either independent or weakly dependent. $\delta_{1t}$ captures the dynamic treatment effect in time period $t$. Following \cite{hahn2017synthetic}, the potential outcomes without the treatment are generated using a factor model for each control unit:
$$
Y_{jt}(0)= Y_{jt} =\lambda_{1j}+F_{1t}+\lambda_{2j} F_{2t}+\epsilon_{jt}, \quad j=2,\dots,J+1,
$$
where $\lambda_{1j}=(j-1) / J, \lambda_{2 j}=(j-1) / J, F_{1t}, F_{2t} \stackrel{iid}{\sim} N(0, 1)$.

\begin{table}[h]
\caption{\small Data generating processes used in simulation studies.}
\centering
\small
\begin{tabular}{|l|l|}
\hline & Weight Specification \\
\hline
DGP 1 & $ \omega^{0} = 1/J\cdot\one_J$ \\
DGP 2 & $ \omega^{0} = (\omega_2, \ldots, \omega_{J+1})^{\top}$, where $\omega_j \sim \text{Unif}(-3/J, 3/J)$ \\
DGP 3 & $ \omega^{0} = (\omega_2, \ldots, \omega_{J+1})^{\top}$, where $\omega_j \sim (\text{Beta}(0.2, 0.2) - 0.5)\times 3/J$ \\
DGP 4 & $ \omega^{0} = (\omega_2, \ldots, \omega_{J/2+1}, \zero_{J/2})^{\top}$, where $\omega_j \sim (\text{Beta}(0.2, 0.2) - 0.5)\times 3/J$  \\
& for $2 \leq j \leq J/2+1$, and shuffle $\omega^{0}$ \\
\hline
\end{tabular}
\end{table}
The generation of weights $\omega^{0}$ is a crucial aspect that differentiates DGPs. Each DGP employs a unique method to specify these weights.
DGP 1 assumes equal weights summing to 1, satisfying all assumptions of synthetic control and providing an intuitive baseline.  
DGP 2 introduces variability by generating weights from a Uniform distribution, adding randomness.  
DGP 3 also incorporates randomness, generating weights from a Beta distribution and rescaling them to have a mean of zero. This approach maintains weights with similar absolute values, blending features of both DGP 1 and DGP 2.
DGP 4 combines multiple strategies, generating half of the weights from a Beta distribution and setting the remaining weights to zero, introducing sparsity to evaluate the strengths and weaknesses of different methods.

\begin{table}[!ht]
\caption{RMSE of Average Treatment Effects (ATE), with the best performance highlighted in yellow and the second-best in green (\textbf{Independent Case}).}
\label{tab:sim_indep}
\centering
\resizebox{0.9\textwidth}{!}{%
\begin{tabular}{@{}llrrrrr llrrrrr@{}}
\toprule
DGP & Method & {$T_0 + 1$} & {$T_0 + 4$} & {$T_0 + 7$} & {$T_0 + 10$} & \quad &
DGP & Method & {$T_0 + 1$} & {$T_0 + 4$} & {$T_0 + 7$} & {$T_0 + 10$}  \\
\midrule
\multirow{7}{*}{DGP 1} 
& Oracle & 1.0169 & 0.5116 & 0.4200 & 0.3423 & &
\multirow{7}{*}{DGP 3} 
& Oracle & 1.0744 & 0.4475 & 0.3516 & 0.3085 \\
& SC & 1.0683 & 0.5452 & 0.4746 & 0.3812 & &
& SC & 1.6817 & 1.0555 & 0.7965 & 0.7138 \\
& Lasso & 1.0745 & 0.5576 & 0.4814 & 0.3846 & &
& Lasso & 1.2035 & 0.5324 & 0.3864 & 0.3513 \\
& Ridge & 1.0143 & \hlf{0.5331} & 0.4537 & 0.3670 & &
& Ridge & 1.1656 & 0.5127 & 0.3784 & 0.3471 \\
& Elastic Net & 1.0357 & 0.5462 & 0.4669 & 0.3782 & &
& Elastic Net & 1.1951 & 0.5238 & 0.3850 & 0.3511 \\
& $L_\infty$ & \hls{0.9829} & \hls{0.5396} & \hlf{0.4399} & \hlf{0.3599} & &
& $L_\infty$ & \hlf{1.1186} & \hlf{0.4708} & \hlf{0.3508} & \hlf{0.3309} \\
& $L_1 + L_\infty$ & \hlf{0.9714} & 0.5466 & \hls{0.4488} & \hls{0.3640} & &
& $L_1 + L_\infty$ & \hls{1.1520} & \hls{0.4813} & \hls{0.3626} & \hls{0.3424} \\
\midrule
\multirow{7}{*}{DGP 2} 
& Oracle & 1.0173 & 0.5109 & 0.4198 & 0.3422 & &
\multirow{7}{*}{DGP 4} 
& Oracle & 1.0751 & 0.4478 & 0.3521 & 0.3090 \\
& SC & 2.0732 & 1.0914 & 0.9826 & 0.9089 & &
& SC & 1.9224 & 1.0431 & 0.7151 & 0.6671 \\
& Lasso & 1.0949 & 0.5502 & 0.4785 & 0.3833 & &
& Lasso & 1.1608 & 0.5086 & \hls{0.3617} & \hls{0.3273} \\
& Ridge & 1.0722 & 0.5461 & 0.4724 & 0.3757 & &
& Ridge & 1.1803 & 0.5194 & 0.3724 & 0.3351 \\
& Elastic Net & 1.0788 & \hls{0.5449} & 0.4750 & 0.3791 & &
& Elastic Net & 1.1628 & \hls{0.5069} & 0.3646 & 0.3278 \\
& $L_\infty$ & \hlf{1.0501} & 0.5502 & \hlf{0.4661} & \hlf{0.3660} & &
& $L_\infty$ & \hls{1.1601} & 0.5259 & 0.3767 & 0.3452 \\
& $L_1 + L_\infty$ & \hls{1.0692} & \hlf{0.5434} & \hls{0.4681} & \hls{0.3720} & &
& $L_1 + L_\infty$ & \hlf{1.1448} & \hlf{0.5024} & \hlf{0.3606} & \hlf{0.3253} \\
\bottomrule
\end{tabular}
}
\end{table}

We conduct all experiments with $T_0=100$, $T_1=10$, $J=30$, and $\delta_{1t} = 3$.
Next, we consider the problem of estimating weights \(\omega\) and treatment effects $\delta_{1t}, t=T_0 + 1,\dots, T_0 + 10$. 
The weights \(\omega\) and the treatment effect $\delta_{1t}$ are obtained by following the procedure described in Section~\ref{subsec:prm}, and we report the individual RMSE based on the results of $B$ replicates of the experiment, computed as 
$
\left(\frac{1}{B}\sum_{b=1}^B \left(\widehat{\delta}_1^{(b)} - \delta_1\right)^2\right)^{1/2},
$
where $\widehat{\delta}_1^{(b)} = \frac{1}{T_1}\sum_{t=T_0+1}^{T_0+T_1}\widehat{\delta}_{1t}^{(b)}$.
Specifically, we set $B=2,000$ in our study. For conciseness, we report only the RMSE for $\widehat{\delta}_1$ for $T_1 = 1, 4, 7$, and $10$ to demonstrate both short-term and long-term performances.

\subsection{Independent Case}\label{subsec:indep}

In our analysis, we allow for the possibility that the error terms are either independent or weakly dependent. 
When assuming independence, both $u_{1t}$ and $\epsilon_{jt}$ follow Gaussian distributions: specifically, $u_{1t} \sim N(0, 1)$ and $\epsilon_{jt} \sim N(0, 2^2)$. 
In this case, the error terms at different time points are uncorrelated, exhibiting no dependence structure.

Table \ref{tab:sim_indep} presents performance results across different data-generating processes, with the best performance highlighted in yellow and the second-best in green. 
In DGP 1, despite aligning with SC's assumptions, SC is still outperformed by our methods, $L_{\infty}$ and $L_1 + L_{\infty}$. Our methods generally achieve the best performances, except being ranked the second at time period $T_0 + 4$, slightly behind the Ridge regression. However, as the time horizon extends, $L_{\infty}$ demonstrates greater robustness, as expected.
In DGP 2, where weights are randomly distributed within the range of $(-3/J, 3/J)$, $L_{\infty}$ shows the best performance. 
Here, the randomly assigned weights lack distinct sparsity or magnitude, allowing $L_{\infty}$ to impose fewer restrictions on the optimization problem (\ref{eq:prob_pen}).

In DGP 3, weights are generated using a ``standardized'' Beta distribution, and under these conditions, $L_{\infty}$ consistently outperforms, with $L_1 + L_{\infty}$ as a close second, even without sparsity. 
This highlights the advantage of our methods over SC in particular.  
DGP 4 introduces sparsity explicitly. In this setting, we expect methods incorporating an $L_1$ norm to outperform those without it. 
Indeed, our proposed method based on $L_1 + L_{\infty}$ performs best, followed by Lasso and elastic net, which rank second and third due to their ability to variable selection capabilities.

\subsection{Dependent Case}\label{subsec:dep}

\begin{table}[htpb]
\centering
\caption{RMSE of Average Treatment Effects (ATE), with the best performance highlighted in yellow and the second-best in green (\textbf{Dependent Case}).}
\label{tab:sim_dep}

\resizebox{0.5\textheight}{!}{%
\begin{tabular}{@{}llrrrrrrrr@{}}
\toprule
 & & \multicolumn{4}{c}{AR(1)} & \multicolumn{4}{c}{ARMA(1, 1)} \\
\cmidrule(lr){3-6} \cmidrule(lr){7-10} 
DGP & Method & {$T_0 + 1$} & {$T_0 + 4$} & {$T_0 + 7$} & {$T_0 + 10$} & {$T_0 + 1$} & {$T_0 + 4$} & {$T_0 + 7$} & {$T_0 + 10$} \\
\midrule
\multirow{7}{*}{DGP 1} 
& Oracle & 1.0128 & 0.5529 & 0.4576 & 0.3760 & 1.0170 & 0.5440 & 0.4489 & 0.3733 \\
& SC & 1.0499 & 0.5974 & 0.5189 & 0.4190 & 1.0451 & 0.5849 & 0.5096 & 0.4142 \\
& Lasso & 1.0469 & 0.6082 & 0.5246 & 0.4235 & 1.0492 & 0.5999 & 0.5190 & 0.4221 \\
& Ridge & 0.9915 & \hlf{0.5792} & 0.4926 & 0.4026 & \hls{0.9909} & \hlf{0.5696} & 0.4868 & \hls{0.4004} \\
& Elastic Net & 1.0080 & \hls{0.5898} & 0.5029 & 0.4100 & 1.0061 & \hls{0.5774} & 0.4956 & 0.4059 \\
& $L_\infty$ & \hls{0.9778} & 0.5917 & \hlf{0.4846} & \hls{0.4002} & 0.9931 & 0.5819 & \hls{0.4802} & 0.4020 \\
& $L_1 + L_\infty$ & \hlf{0.9573} & 0.5947 & \hls{0.4848} & \hlf{0.3991} & \hlf{0.9646} & 0.5842 & \hlf{0.4783} & \hlf{0.3951} \\
\midrule
\multirow{7}{*}{DGP 2} 
& Oracle & 1.0123 & 0.5532 & 0.4577 & 0.3761 & 1.0147 & 0.5460 & 0.4492 & 0.3727 \\
& SC & 2.0598 & 1.1394 & 1.0200 & 0.9345 & 2.0462 & 1.1339 & 1.0085 & 0.9195 \\
& Lasso & 1.0715 & 0.6030 & 0.5211 & 0.4197 & 1.0637 & \hls{0.5846} & 0.5078 & 0.4145 \\
& Ridge & \hls{1.0535} & \hlf{0.5957} & 0.5098 & 0.4088 & 1.0594 & 0.5873 & 0.5012 & 0.4082 \\
& Elastic Net & 1.0593 & \hls{0.5977} & 0.5160 & 0.4139 & 1.0567 & \hlf{0.5842} & 0.5045 & 0.4096 \\
& $L_\infty$ & \hlf{1.0514} & 0.6052 & \hlf{0.5043} & \hlf{0.4031} & \hlf{1.0375} & 0.5924 & \hls{0.4992} & \hls{0.4027} \\
& $L_1 + L_\infty$ & 1.0542 & 0.5980 & \hls{0.5076} & \hls{0.4065} & \hls{1.0477} & 0.5966 & \hlf{0.4991} & \hlf{0.4017} \\
\midrule
\multirow{7}{*}{DGP 3} 
& Oracle & 1.0823 & 0.4898 & 0.3890 & 0.3438 & 1.0866 & 0.4814 & 0.3842 & 0.3421 \\
& SC & 1.7034 & 1.1043 & 0.8397 & 0.7437 & 1.6858 & 1.0800 & 0.8382 & 0.7407 \\
& Lasso & 1.1967 & 0.5707 & 0.4235 & 0.3881 & 1.1996 & 0.5692 & 0.4276 & 0.3900 \\
& Ridge & 1.1617 & 0.5517 & 0.4140 & 0.3842 & 1.1673 & 0.5483 & 0.4140 & 0.3837 \\
& Elastic Net & 1.1875 & 0.5612 & 0.4207 & 0.3868 & 1.1833 & 0.5580 & 0.4218 & 0.3871 \\
& $L_\infty$ & \hlf{1.1149} & \hlf{0.5038} & \hlf{0.3843} & \hlf{0.3662} & \hlf{1.1349} & \hlf{0.5012} & \hlf{0.3842} & \hlf{0.3662} \\
& $L_1 + L_\infty$ & \hls{1.1396} & \hls{0.5257} & \hls{0.4011} & \hls{0.3788} & \hls{1.1499} & \hls{0.5200} & \hls{0.3985} & \hls{0.3769} \\
\midrule
\multirow{7}{*}{DGP 4} 
& Oracle & 1.0824 & 0.4901 & 0.3877 & 0.3424 & 1.0091 & 0.4713 & 0.3196 & 0.2693 \\
& SC & 1.9242 & 1.0854 & 0.7470 & 0.6873 & 1.4466 & 0.8491 & 0.8322 & 0.7396 \\
& Lasso & 1.1624 & \hls{0.5406} & \hls{0.3956} & \hls{0.3603} & 0.9868 & 0.4899 & 0.3903 & 0.3586 \\
& Ridge & 1.1789 & 0.5547 & 0.4066 & 0.3707 & 0.9797 & \hls{0.4852} & 0.3908 & 0.3531 \\
& Elastic Net & 1.1625 & 0.5427 & 0.3974 & 0.3620 & 0.9886 & 0.4874 & \hls{0.3867} & 0.3527 \\
& $L_\infty$ & \hls{1.1506} & 0.5593 & 0.4109 & 0.3778 & \hlf{0.9441} & 0.4961 & 0.3979 & \hls{0.3521} \\
& $L_1 + L_\infty$ & \hlf{1.1423} & \hlf{0.5369} & \hlf{0.3928} & \hlf{0.3583} & \hls{0.9787} & \hlf{0.4809} & \hlf{0.3784} & \hlf{0.3470} \\
\bottomrule
\end{tabular}
}
\end{table}

We consider two types of dependent processes: an AR(1) process and an ARMA(1,1) process given by:

\begin{enumerate}
    \item [Case 1:] \textbf{AR}(1)    $u_{1t} = \rho u_{1,t-1} + \zeta_{1t}$ and $\epsilon_{jt} = \rho \epsilon_{j,t-1} + \zeta_{jt}$, where $\rho = 0.1$ is the autoregressive coefficient. 
Here, $\zeta_{jt}$ represents white noise error terms, each following a standard Gaussian distribution.

\item [Case 2:] \textbf{ARMA}(1,1)     $u_{1t} = \rho u_{1,t-1} + \theta \zeta_{1,t-1} + \zeta_{1t}$ and $\epsilon_{jt} = \rho \epsilon_{j,t-1} + \theta \zeta_{j,t-1} + \zeta_{jt}$, where $\rho = 0.1$ is the autoregressive coefficient, $\theta = 0.1$ is the moving average coefficient, and $\zeta_{1t}$ and $\zeta_{jt}$ are white noise error terms following a standard Gaussian distribution.
\end{enumerate}

\noindent To ensure comparability with the results in Section~\ref{subsec:dep}, we multiply $\epsilon_{jt}$ by a factor of 2 to match the scale used there. Table \ref{tab:sim_dep} presents performance results. In DGP 1, the results exhibit patterns similar to those observed in the independent case. SC remains non-competitive. More importantly, as the time horizon extends, $L_1 + L_{\infty}$--rather than $L_{\infty}$--demonstrates the best performance.
In DGP 2, where weights are randomly distributed within the range $(-3/J, 3/J)$, $L_{\infty}$ performs best under the AR(1) scenario, while $L_1 + L_{\infty}$ shows superior performance with ARMA(1,1). This suggests that the inclusion of an $L_1$ norm can be beneficial when a moving average component is present.

In DGP 3, the dependency structure does not alter the relative performance ranking, with $L_{\infty}$ and $L_1 + L_{\infty}$ consistently outperforming other methods. 
In DGP 4, $L_1 + L_{\infty}$, which allows for a denser weighting scheme, achieves the best performance. While it is challenging to rank the remaining methods precisely, we observe that the $L_1$ norm is advantageous for handling sparsity when present.

\section{Real Data}\label{sec:real_data}

\subsection{California Tobacco}

In this analysis, we revisit the impact of California's Proposition 99, a landmark public health policy aimed at reducing tobacco use through higher taxation and comprehensive anti-smoking initiatives. Since its introduction, this policy has been extensively studied, including in \cite{abadie_synthetic_2010}. Here, we compare our proposed methodology with the established methods discussed in Section \ref{sec:sim}.

Proposition 99, passed in November 1988 and enacted in January 1989, introduced a 25-cent per-pack tax on cigarettes, directing the revenue toward various health programs. 
The data spans from 1970 to 2000, allowing us to observe smoking trends before and after the policy's implementation.
The dataset includes annual state-level panel data on cigarette sales across 38 states, selected to exclude those that enacted significant tobacco control measures or major tax hikes during the study period \citep{abadie_synthetic_2010}. 
The response variable is annual per capita cigarette sales, measured in packs per capita at the state level.

We apply the same parameter tuning strategies used in the simulation studies. 
Using the pre-treatment data, we set $K = T_0$, implementing a leave-one-out cross-validation procedure to optimize $\alpha$ and $\lambda$ by minimizing the RMSE. With these optimized parameters, we then determined the weights for each method.

\begin{figure}[ht]
    \centering
    \includegraphics[width=0.8\textwidth]{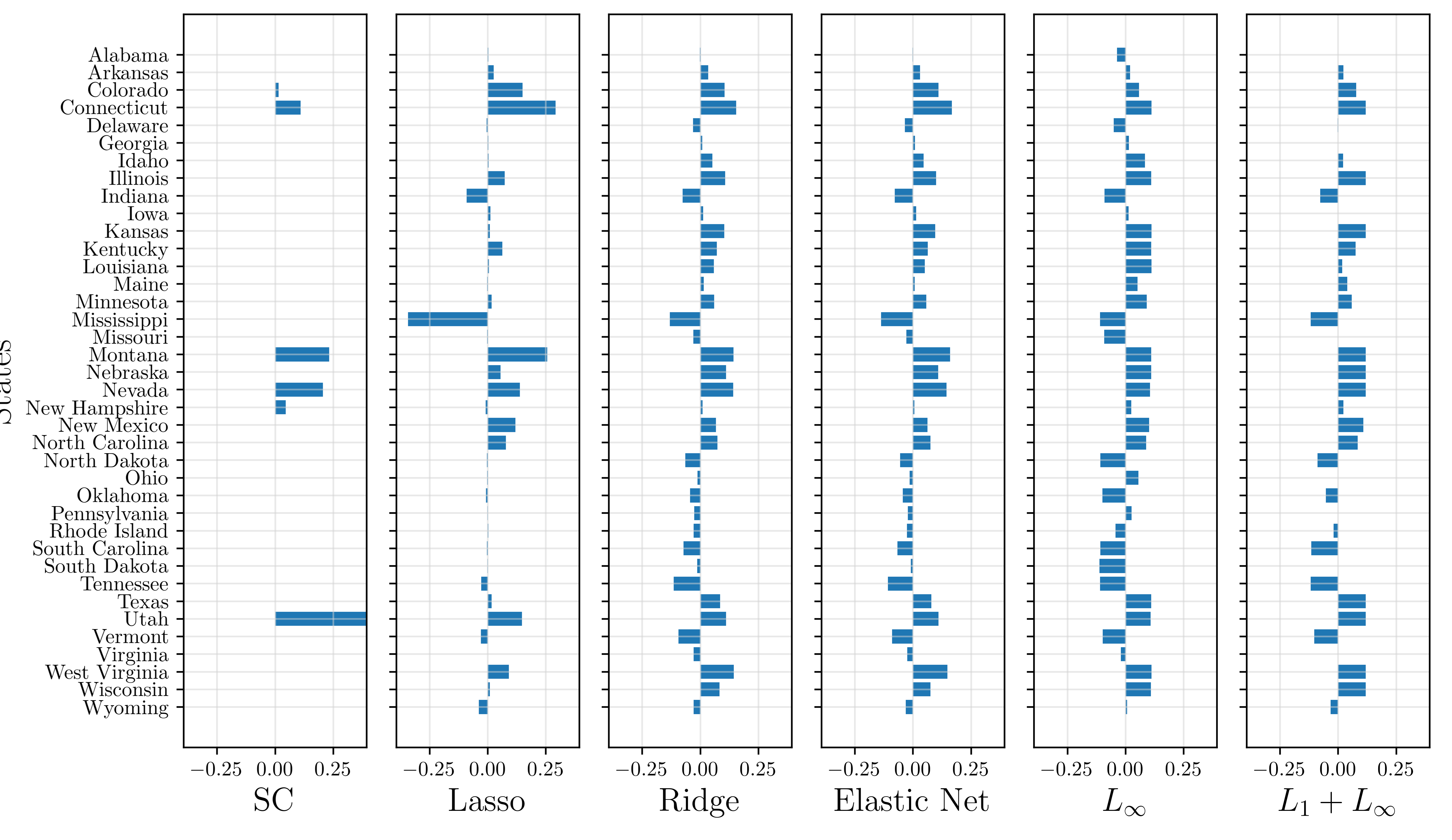}
    
    \caption{\small State Weights}
    \label{fig:tobacco_weights}
\end{figure}

Figure \ref{fig:tobacco_weights} displays the results: SC assigns the majority of its weights to six states: Utah, New Hampshire, Nevada, Montana, Connecticut, and Colorado. 
While some of these states are geographically close to California, others are much farther from the West Coast and differ in several important ways, making the interpretation of the weight allocation challenging. 
We also observe potential instability in SC due to the high sparsity of the weights. 
In contrast, the other methods distribute weights across a broader range of states. Lasso regression, though still sparse, exhibits slightly more distributed than SC, while ridge and elastic net, as well as $L_{\infty}$ and $L_1 + L_{\infty}$, allocate weights more evenly. 
Notably, the similarity between ridge and elastic net, as well as $L_{\infty}$ and $L_1 + L_{\infty}$, can be attributed to the small values of $\alpha$ selected in this study.

\begin{figure}[ht]
    \centering
    \includegraphics[width=0.6\textwidth]{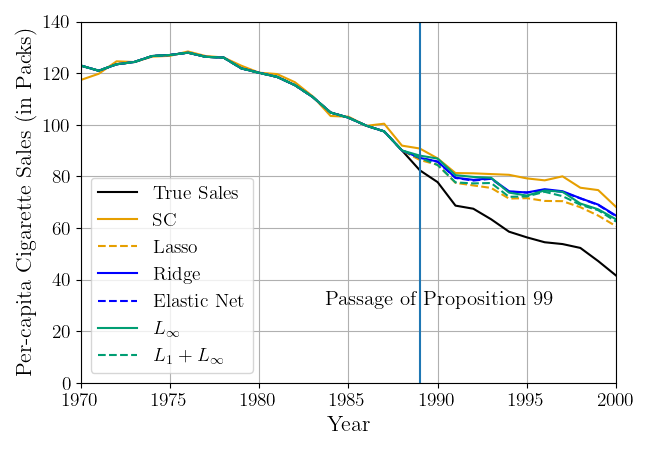}
    
    \caption{\small Sales Trends of Synthetic California}
    \label{fig:tobacco}
\end{figure}

\begin{figure}[ht]
    \centering
    \includegraphics[width=0.9\textwidth]{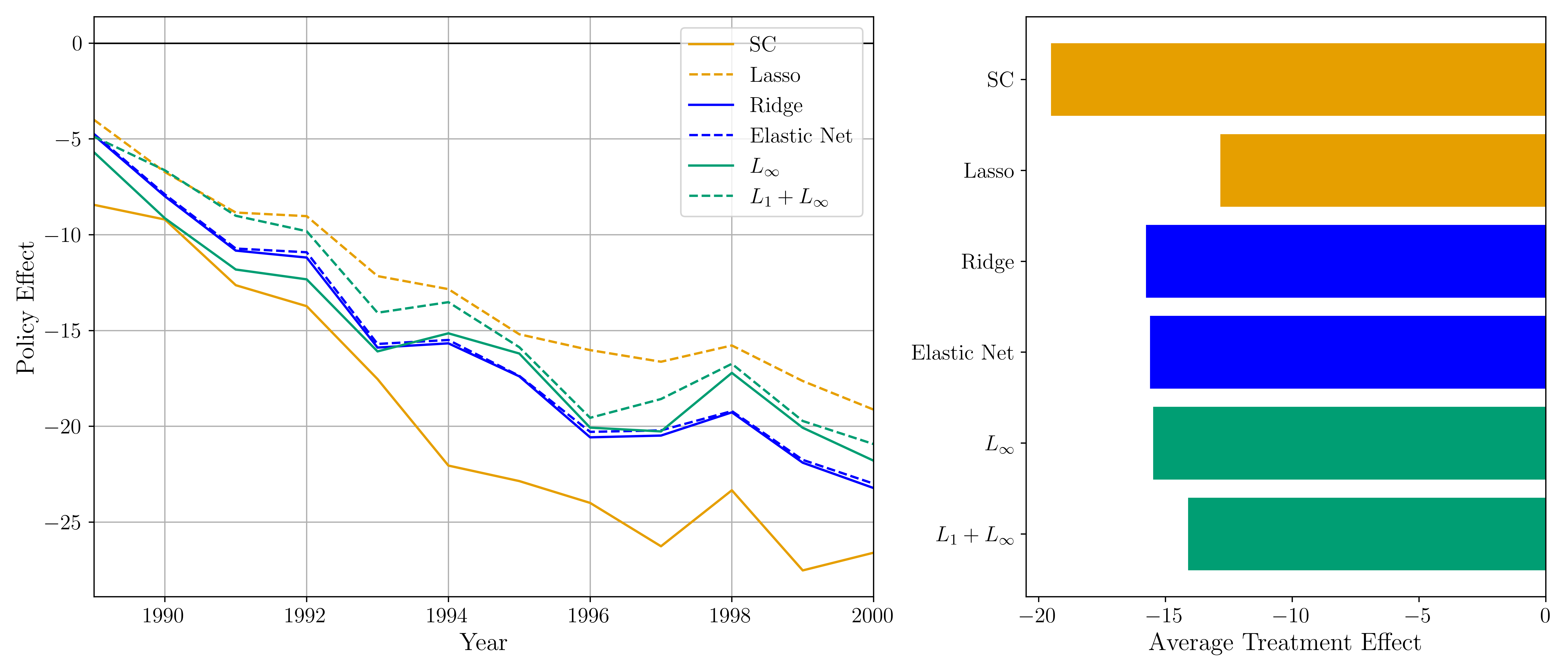}
    
    \caption{\small Left Panel: The Estimates of Policy Effects; Right Panel: The Average Treatment Effect}
    \label{fig:tobacco_trt}
\end{figure}

Figure~\ref{fig:tobacco} displays the performance of the synthetic Californias. 
In the left panel, we observe that all methods align well with the true data before the treatment begins. 
After the passage of Proposition 99, other methods exhibit similar prediction trends. 
In the right panel, the treatment effects are consistently negative, supporting the conclusion that Proposition 99 effectively reduced cigarette consumption. 
However, we note that SC appears to overestimate the effect, deviating noticeably from the trends observed in other methods, while the Lasso method underestimates the effect. On the other hand, the estimated average treatment effect is similar between our proposed methods and the existing ones based on alternative regularization methods. This is not surprising since, in this canonical example for synthetic control methods, all these methods achieve desirable matches between the synthetic control and the treated unit. More importantly, these methods allocate the weights more evenly in this context, avoiding the substantial instability in estimates for the sparse weighting schemes such as those for the conventional SC and the Lasso-based SC methods. This contrast also highlights the inherently added advantage of our methods as they maintain the same level of the desirable pre-treatment matches while the resulting denser weighting schemes can avoid the potential impact of reliance on a few control units and enhance robustness.

\subsection{Regulation SHO}

In this section, we examine the effect of a regulatory policy in the stock market, a highly volatile environment where conventional synthetic control (SC) methods relying on a few control units are more susceptible to shocks and biased estimates. This real-world example effectively illustrates the advantages of our proposed methods and the benefits of employing a denser weighting scheme in constructing the synthetic control.

Specifically, we analyze the impact of the Regulation SHO Pilot Program on the stock market. This program, initiated by the U.S. Securities and Exchange Commission (SEC), was designed to assess the effects of removing short-sale price tests, such as the uptick rule, on market quality, trading behavior, and volatility. The program included approximately one-third of the firms listed in the Russell 3000 index, which were randomly assigned to three distinct categories based on when short-sale price tests were suspended: \textbf{Category A}: No price test at any time. \textbf{Category B}: No price test from 4:15 PM ET until the next day's market opening. \textbf{Category C} No price test from market close until the following market open. The pilot program ran from May 2, 2005, to April 28, 2006.

\begin{figure}[!h]
    \centering
    \includegraphics[width=0.8\textwidth]{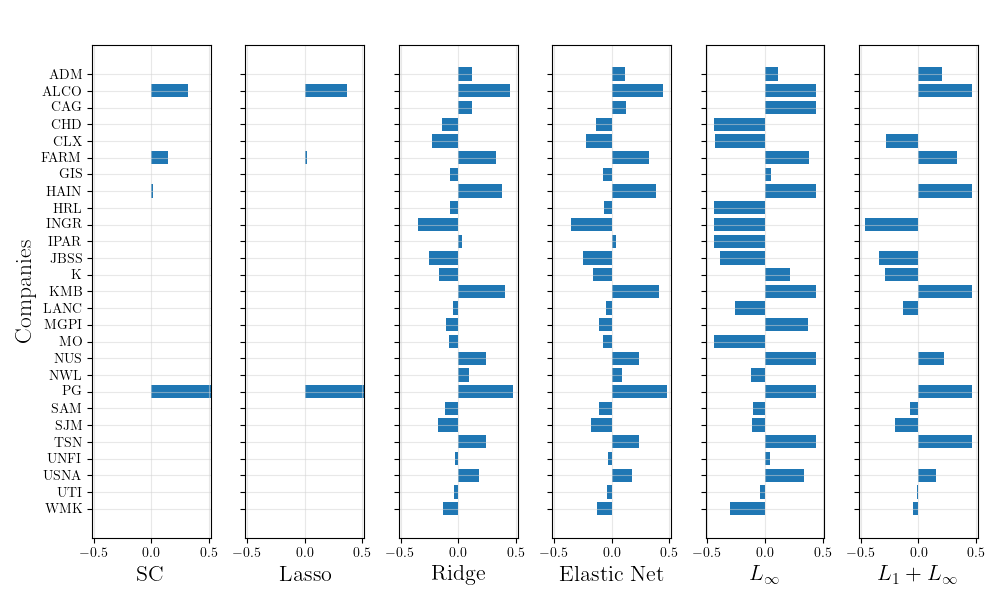}
    \caption{\small Company Weights}
    \label{fig:sho_weights}
\end{figure}

We apply our proposed methods to evaluate the impact of the Regulation SHO Program on weekly average stock prices. For our analysis, we obtain historical data from Yahoo Finance, identifying 761 companies with available data from January 1, 2005, to May 31, 2005. To illustrate our approach, we focus on the \emph{Consumer Defensive} sector, often considered less volatile than sectors like technology or finance, meaning that changes in stock price due to Regulation SHO are less likely to be confounded by external macroeconomic shocks. We also consider the policy impact on the stock prices of \emph{Costco Wholesale Corporation} since, as a large-cap publicly traded company, Costco has high liquidity and significant trading volume. This makes it an ideal candidate for studying the effects of removing short-sale price tests, as any distortions caused by regulatory changes would be more observable in such actively traded stocks. We use 27 unaffected companies as the control group.

We employ the leave-one-out cross-validation approach to optimize key model parameters, specifically $\alpha$ and $\lambda$. Figure \ref{fig:sho_weights} presents the results of this approach. In this context, we find that both the traditional synthetic control method and the Lasso-based approaches tend to concentrate weights on a small subset of companies (four and three, respectively), making the estimates overly dependent on a few firms. In contrast, other methods distribute weights more evenly across the control units. The methods, in particular, achieve a balance by maintaining sparsity while ensuring well-regularized weight magnitudes. Furthermore, our approaches adaptively assign slightly larger weights to certain control units, with the maximum weight constrained by the tuning parameters.

\begin{figure}[htb]
    \centering
    \includegraphics[width=0.6\textwidth]{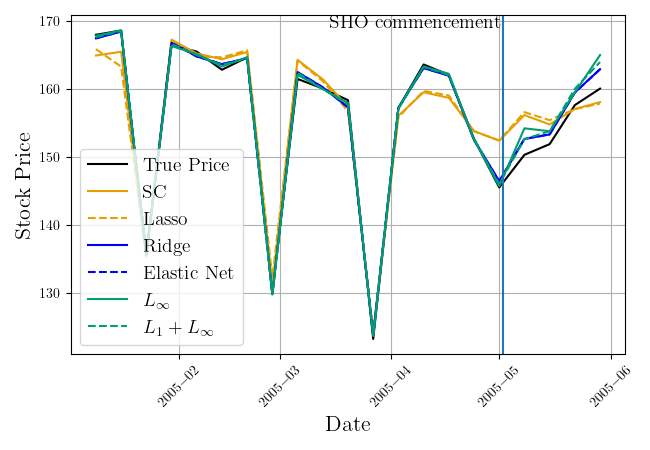}
    
    \caption{\small Stock Price of Synthetic ``Costco''}
    \label{fig:sho}
\end{figure}

In Figure \ref{fig:sho}, we can see that prior to the SHO intervention, all estimation methods (except the conventional and Lasso-based SC methods) closely track the true stock price trajectory, demonstrating that the models effectively capture the pre-treatment dynamics. However, after policy implementation, the models suggest an upward trend in counterfactual stock prices at a level significantly higher than what the actual prices indicate. This result suggests the presence of potentially negative policy effects on stock prices.

\begin{figure}[htb]
    \centering
    \includegraphics[width=0.85\textwidth]{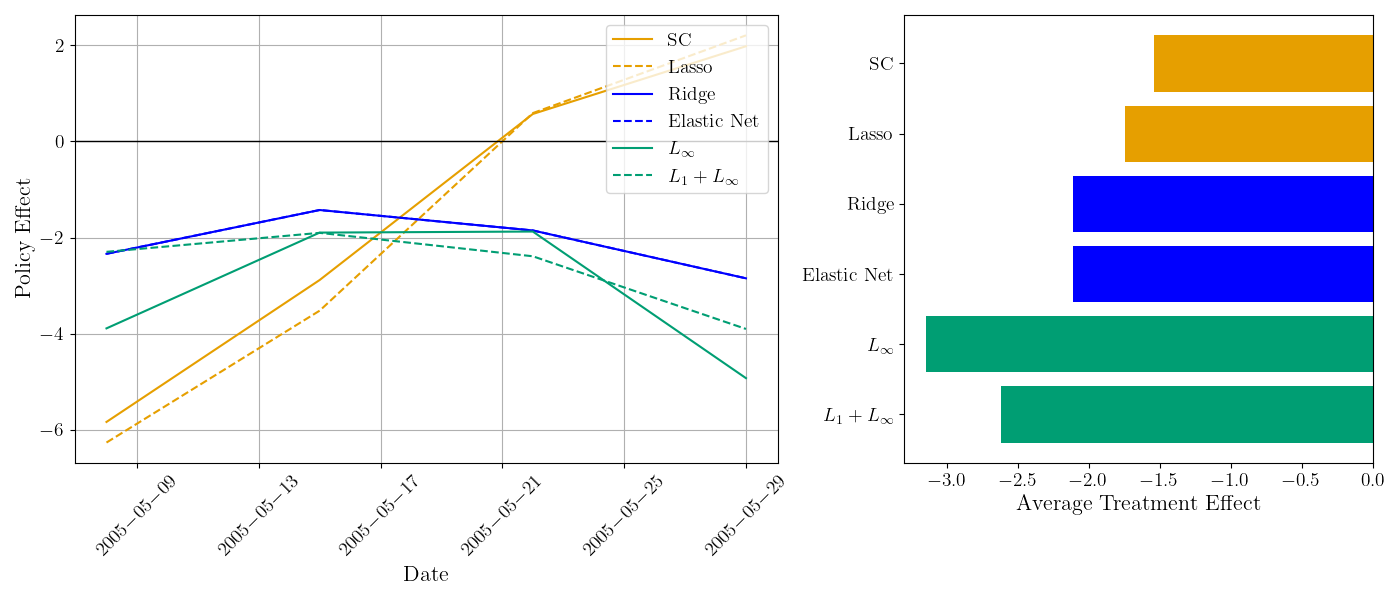}
    
    \caption{\small Left Panel: The Estimates of Policy Effects; Right Panel: The Average Treatment Effect}
    \label{fig:sho_trt}
    
\end{figure}

Figures~\ref{fig:sho} and \ref{fig:sho_trt} also plot the actual policy effect estimated by these different methods. This graph represents the difference between the estimated counterfactual prices and the actual observed prices. As shown, all methods indeed begin to detect a negative policy effect shortly after the SHO commencement in early May. 

However, there are important differences in the dynamic treatment effects and, hence, the average treatment effects across different models. Both the conventional and the Lasso-based results suggest that the initial policy effect is gradually declining over time and eventually disappears and even turns positive at the end of the study period. On the other hand, the methods relying on regularization methods that generate denser weights indicate that the policy effect persists. More importantly, both the $L_1 + L_\infty$ and $L_\infty$ methods (dashed green line and solid green line, respectively) also suggest potentially larger negative policy effects. 

The significant contrast between the conventional/Lasso-based synthetic control and alternative approaches aligns with the observations from Figure \ref{fig:sho_weights}, where the conventional and Lasso-based synthetic control methods rely heavily on a small number of control units; and as expected, it is likely that in a responsive financial market, alternative shocks to the firms with concentrated weights affect their prices after the policy and lead to biased estimates of the policy effects. While we do not have the ground truth for the treatment effect in a real-life sceniario like this, our simulation results suggest the better performances of our approaches. It is therefore more likely that the more balanced weight distribution in methods like $L_1 + L_\infty$ results in better control over the estimation, offering a potentially more reliable interpretation of the policy's effect.

\section{Discussion}\label{sec:discussion}

The SC technique is a widely recognized and effective method in causal inference \citep{athey_state_2017}; however, it is often accompanied by several unnecessary constraints. In this paper, we propose to relax these constraints and employ the $L_{\infty}$ norm as a potential alternative to the more commonly employed $L_1$ and $L_2$ norms in penalized regression.
Leveraging the theoretical framework established by \cite{li_statistical_2020}, we derive the distributional properties of our weight estimator $\sol$, as well as the ATE estimator $\widehat{\delta}_1$. This enables us to provide a novel approach for constructing flexible and effective estimators in the context of causal analysis.
Extensive simulation studies and real-world data examples showcase the strengths of our proposed method. Over the long term, our approach demonstrates robustness in making accurate predictions. 
We acknowledge that the performance of our method is influenced by the underlying true weights. When the pattern of the true weights ($\omega^0$) aligns well with the strengths of our method, it significantly outperforms competing approaches. 
Even in more general cases, our method remains competitive and performs comparably to other established methods. 
In the future, there are at least two directions worth exploring. It would be beneficial to develop theories for short panels where $T_0 > J$ is violated. We aim to further develop the inference theory based on the existing framework, specifically focusing on constructing confidence intervals.


\bibliographystyle{abbrvnat}

\begin{thebibliography}{25}
\providecommand{\natexlab}[1]{#1}
\providecommand{\url}[1]{\texttt{#1}}
\expandafter\ifx\csname urlstyle\endcsname\relax
  \providecommand{\doi}[1]{doi: #1}\else
  \providecommand{\doi}{doi: \begingroup \urlstyle{rm}\Url}\fi

\bibitem[Abadie et~al.(2010)Abadie, Diamond, and
  Hainmueller]{abadie_synthetic_2010}
A.~Abadie, A.~Diamond, and J.~Hainmueller.
\newblock Synthetic control methods for comparative case studies: estimating
  the effect of california’s tobacco control program.
\newblock \emph{Journal of the American Statistical Association}, 105\penalty0
  (490):\penalty0 493--505, 2010.
\newblock ISSN 0162-1459, 1537-274X.
\newblock \doi{10.1198/jasa.2009.ap08746}.

\bibitem[Abadie et~al.(2015)Abadie, Diamond, and
  Hainmueller]{abadie_comparative_2015}
A.~Abadie, A.~Diamond, and J.~Hainmueller.
\newblock Comparative politics and the synthetic control method: Comparative
  politics and the synthetic control method.
\newblock \emph{American Journal of Political Science}, 59\penalty0
  (2):\penalty0 495--510, 2015.
\newblock ISSN 00925853.
\newblock \doi{10.1111/ajps.12116}.

\bibitem[Andersen et~al.(2003)Andersen, Roos, and
  Terlaky]{andersen2003implementing}
E.~D. Andersen, C.~Roos, and T.~Terlaky.
\newblock On implementing a primal-dual interior-point method for conic
  quadratic optimization.
\newblock \emph{Mathematical Programming}, 95:\penalty0 249--277, 2003.

\bibitem[Arkhangelsky et~al.(2021)Arkhangelsky, Athey, Hirshberg, Imbens, and
  Wager]{arkhangelsky2021synthetic}
D.~Arkhangelsky, S.~Athey, D.~A. Hirshberg, G.~W. Imbens, and S.~Wager.
\newblock Synthetic difference-in-differences.
\newblock \emph{American Economic Review}, 111\penalty0 (12):\penalty0
  4088--4118, 2021.

\bibitem[Athey and Imbens(2017)]{athey_state_2017}
S.~Athey and G.~W. Imbens.
\newblock The {State} of {Applied} {Econometrics}: {Causality} and {Policy}
  {Evaluation}.
\newblock \emph{Journal of Economic Perspectives}, 31\penalty0 (2):\penalty0
  3--32, 2017.
\newblock ISSN 0895-3309.
\newblock \doi{10.1257/jep.31.2.3}.

\bibitem[Athey et~al.(2021)Athey, Bayati, Doudchenko, Imbens, and
  Khosravi]{athey2021matrix}
S.~Athey, M.~Bayati, N.~Doudchenko, G.~Imbens, and K.~Khosravi.
\newblock Matrix completion methods for causal panel data models.
\newblock \emph{Journal of the American Statistical Association}, 116\penalty0
  (536):\penalty0 1716--1730, 2021.

\bibitem[Billmeier and Nannicini(2013)]{billmeier2013assessing}
A.~Billmeier and T.~Nannicini.
\newblock Assessing economic liberalization episodes: A synthetic control
  approach.
\newblock \emph{Review of Economics and Statistics}, 95\penalty0 (3):\penalty0
  983--1001, 2013.

\bibitem[Doudchenko and Imbens(2016)]{doudchenko_balancing_2016}
N.~Doudchenko and G.~Imbens.
\newblock Balancing, {Regression}, {Difference}-{In}-{Differences} and
  {Synthetic} {Control} {Methods}: {A} {Synthesis}.
\newblock Technical Report w22791, National Bureau of Economic Research,
  Cambridge, MA, 2016.

\bibitem[Doukhan and Louhichi(1999)]{doukhan1999new}
P.~Doukhan and S.~Louhichi.
\newblock A new weak dependence condition and applications to moment
  inequalities.
\newblock \emph{Stochastic Processes and their Applications}, 84\penalty0
  (2):\penalty0 313--342, 1999.
\newblock ISSN 0304-4149.
\newblock \doi{https://doi.org/10.1016/S0304-4149(99)00055-1}.

\bibitem[Friedman et~al.(2010)Friedman, Tibshirani, and
  Hastie]{friedman2010glmnet}
J.~Friedman, R.~Tibshirani, and T.~Hastie.
\newblock Regularization paths for generalized linear models via coordinate
  descent.
\newblock \emph{Journal of Statistical Software}, 33\penalty0 (1):\penalty0
  1--22, 2010.
\newblock \doi{10.18637/jss.v033.i01}.

\bibitem[Fu and Knight(2000)]{fu_2000_asymptotics}
W.~Fu and K.~Knight.
\newblock {Asymptotics for lasso-type estimators}.
\newblock \emph{The Annals of Statistics}, 28\penalty0 (5):\penalty0 1356 --
  1378, 2000.
\newblock \doi{10.1214/aos/1015957397}.

\bibitem[Hahn and Shi(2017)]{hahn2017synthetic}
J.~Hahn and R.~Shi.
\newblock Synthetic control and inference.
\newblock \emph{Econometrics}, 5\penalty0 (4):\penalty0 n.p., 2017.
\newblock ISSN 2225-1146.

\bibitem[Lei and Candès(2021)]{lei2021conformal}
L.~Lei and E.~J. Candès.
\newblock Conformal inference of counterfactuals and individual treatment
  effects.
\newblock \emph{Journal of the Royal Statistical Society Series B: Statistical
  Methodology}, 83\penalty0 (5):\penalty0 911--938, 10 2021.
\newblock ISSN 1369-7412.
\newblock \doi{10.1111/rssb.12445}.
\newblock URL \url{https://doi.org/10.1111/rssb.12445}.

\bibitem[Li(2020)]{li_statistical_2020}
K.~T. Li.
\newblock Statistical {Inference} for {Average} {Treatment} {Effects}
  {Estimated} by {Synthetic} {Control} {Methods}.
\newblock \emph{Journal of the American Statistical Association}, 115\penalty0
  (532):\penalty0 2068--2083, 2020.
\newblock ISSN 0162-1459, 1537-274X.
\newblock \doi{10.1080/01621459.2019.1686986}.

\bibitem[Liao et~al.(2023)Liao, Ma, Neuhierl, and Shi]{liao2023economic}
Y.~Liao, X.~Ma, A.~Neuhierl, and Z.~Shi.
\newblock Economic forecasts using many noises.
\newblock \emph{arXiv preprint arXiv:2312.05593}, 2023.

\bibitem[McClelland and Gault(2017)]{mcclelland2017synthetic}
R.~McClelland and S.~Gault.
\newblock The synthetic control method as a tool to understand state policy.
\newblock \emph{Washington, DC: Urban-Brookings Tax Policy Center}, 2017.

\bibitem[Rehkopf and Basu(2018)]{rehkopf2018new}
D.~H. Rehkopf and S.~Basu.
\newblock A new tool for case studies in epidemiology—the synthetic control
  method.
\newblock \emph{Epidemiology}, 29\penalty0 (4):\penalty0 503--505, 2018.

\bibitem[Rubin(1974)]{rubin_1974}
D.~B. Rubin.
\newblock Estimating causal effects of treatments in randomized and
  nonrandomized studies.
\newblock \emph{Journal of educational Psychology}, 66\penalty0 (5):\penalty0
  688, 1974.

\bibitem[Shen et~al.(2022)Shen, Wan, Cai, and Song]{shen2022heterogeneous}
Y.~Shen, R.~Wan, H.~Cai, and R.~Song.
\newblock Heterogeneous synthetic learner for panel data, 2022.

\bibitem[Shi et~al.(2025)Shi, Su, and Xie]{shi2025ℓ}
Z.~Shi, L.~Su, and T.~Xie.
\newblock $l_2$-relaxation: With applications to forecast combination and
  portfolio analysis.
\newblock \emph{Review of Economics and Statistics}, 107\penalty0 (2):\penalty0
  523--538, 2025.

\bibitem[Stefano and Mellace(2024)]{distefano2024inclusive}
R.~D. Stefano and G.~Mellace.
\newblock The inclusive synthetic control method, 2024.
\newblock URL \url{https://arxiv.org/abs/2403.17624}.

\bibitem[Tirunillai and Tellis(2017)]{tirunillai2017does}
S.~Tirunillai and G.~J. Tellis.
\newblock Does offline tv advertising affect online chatter? quasi-experimental
  analysis using synthetic control.
\newblock \emph{Marketing Science}, 36\penalty0 (6):\penalty0 862--878, 2017.

\bibitem[T{\"u}t{\"u}nc{\"u} et~al.(2003)T{\"u}t{\"u}nc{\"u}, Toh, and
  Todd]{tutuncu2003solving}
R.~H. T{\"u}t{\"u}nc{\"u}, K.-C. Toh, and M.~J. Todd.
\newblock Solving semidefinite-quadratic-linear programs using sdpt3.
\newblock \emph{Mathematical programming}, 95:\penalty0 189--217, 2003.

\bibitem[Wright(1997)]{wright1997primal}
S.~J. Wright.
\newblock \emph{Primal-Dual Interior-Point Methods}.
\newblock Society for Industrial and Applied Mathematics (SIAM), Philadelphia,
  PA, 1997.
\newblock ISBN 978-0-89871-382-6.
\newblock \doi{10.1137/1.9781611971453}.

\bibitem[Zou and Hastie(2005)]{zou_regularization_2005}
H.~Zou and T.~Hastie.
\newblock Regularization and {Variable} {Selection} {Via} the {Elastic} {Net}.
\newblock \emph{Journal of the Royal Statistical Society Series B: Statistical
  Methodology}, 67\penalty0 (2):\penalty0 301--320, 2005.
\newblock ISSN 1369-7412.
\newblock \doi{10.1111/j.1467-9868.2005.00503.x}.

\end{thebibliography}

\end{document}